\documentclass[12pt]{article}

\usepackage{mathrsfs}
\usepackage[T1]{fontenc}
\usepackage{mathpazo}
\usepackage{setspace}
\usepackage{amsfonts}
\usepackage{amssymb}
\usepackage{amsmath}
\usepackage{esint}                   
\usepackage{epsfig}
\usepackage{latexsym}
\usepackage{color}
\usepackage{graphicx}
\usepackage{hyperref}
\usepackage{nicefrac}
\usepackage[latin1]{inputenc}
\usepackage{pstricks}
\usepackage{slashed}
\usepackage{multirow}
\usepackage{ulem}
\usepackage{comment}
\usepackage[nosort]{cite}
\usepackage{datetime}
\usepackage{tikz-cd}
\usepackage{float}

\usepackage{relsize}

\usepackage{tikz}
\usepackage{capt-of}
\usetikzlibrary{decorations.markings}

\usepackage[greek,english]{babel}
\languageattribute{greek}{polutoniko}

\usepackage[vcentermath]{youngtab}

\def\hybrid{\topmargin -30pt    \oddsidemargin 0pt 
        \headheight 0pt \headsep 0pt
        \textwidth 6.25in       
        \textheight 9.5in       
        \marginparwidth .875in
        \parskip 5pt plus 1pt   \jot = 1.5ex}

\hybrid

\def\baselinestretch{1.2}

\catcode`\@=11

\def\marginnote#1{}
\newcount\hour
\newcount\minute
\newtoks\amorpm
\hour=\time\divide\hour by60
\minute=\time{\multiply\hour by60 \global\advance\minute by-\hour}
\edef\standardtime{{\ifnum\hour<12 \global\amorpm={am}%
        \else\global\amorpm={pm}\advance\hour by-12 \fi
        \ifnum\hour=0 \hour=12 \fi
        \number\hour:\ifnum\minute<10 0\fi\number\minute\the\amorpm}}
\edef\militarytime{\number\hour:\ifnum\minute<10 0\fi\number\minute}

\def\draftlabel#1{{\@bsphack\if@filesw {\let\thepage\relax
   \xdef\@gtempa{\write\@auxout{\string
      \newlabel{#1}{{\@currentlabel}{\thepage}}}}}\@gtempa
   \if@nobreak \ifvmode\nobreak\fi\fi\fi\@esphack}
        \gdef\@eqnlabel{#1}}
\def\@eqnlabel{}
\def\@vacuum{}
\def\draftmarginnote#1{\marginpar{\raggedright\scriptsize\tt#1}}

\def\draft{\oddsidemargin -.5truein
        \def\@oddfoot{\sl preliminary draft \hfil
        \rm\thepage\hfil\sl\today\quad\militarytime}
        \let\@evenfoot\@oddfoot \overfullrule 3pt
        \let\label=\draftlabel
        \let\marginnote=\draftmarginnote
   \def\@eqnnum{(\theequation)\rlap{\kern\marginparsep\tt\@eqnlabel}%
\global\let\@eqnlabel\@vacuum}  }

\def\draft2{
        \def\@oddfoot{\sl preliminary draft \hfil
        \rm\thepage\hfil\sl\today\quad\militarytime}
        \let\@evenfoot\@oddfoot \overfullrule 3pt
        \let\marginnote=\draftmarginnote
   \def\@eqnnum{(\theequation)\rlap{\kern\marginparsep\tt\@eqnlabel}%
\global\let\@eqnlabel\@vacuum}  }

\def\preprint{\twocolumn\sloppy\flushbottom\parindent 2em
        \leftmargini 2em\leftmarginv .5em\leftmarginvi .5em
        \oddsidemargin -.5in    \evensidemargin -.5in
        \columnsep .4in \footheight 0pt
        \textwidth 10.in        \topmargin  -.4in
        \headheight 12pt \topskip .4in
        \textheight 6.9in \footskip 0pt
        \def\@oddhead{\thepage\hfil\addtocounter{page}{1}\thepage}
        \let\@evenhead\@oddhead \def\@oddfoot{} \def\@evenfoot{} }

\def\numberbysection{\@addtoreset{equation}{section}
        \def\theequation{\thesection.\arabic{equation}}}

\def\underline#1{\relax\ifmmode\@@underline#1\else
        $\@@underline{\hbox{#1}}$\relax\fi}

\def\titlepage{\@restonecolfalse\if@twocolumn\@restonecoltrue\onecolumn
     \else \newpage \fi \thispagestyle{empty}\c@page\z@
        \def\thefootnote{\fnsymbol{footnote}} }

\def\endtitlepage{\if@restonecol\twocolumn \else \newpage \fi
        \def\thefootnote{\arabic{footnote}}
        \setcounter{footnote}{0}}  

\catcode`@=12
\relax

\def\figcap{\section*{Figure Captions\markboth
        {FIGURECAPTIONS}{FIGURECAPTIONS}}\list
        {Figure \arabic{enumi}:\hfill}{\settowidth\labelwidth{Figure
999:}
        \leftmargin\labelwidth
        \advance\leftmargin\labelsep\usecounter{enumi}}}
 \relax
\def\tablecap{\section*{Table Captions\markboth
        {TABLECAPTIONS}{TABLECAPTIONS}}\list
        {Table \arabic{enumi}:\hfill}{\settowidth\labelwidth{Table
999:}
        \leftmargin\labelwidth
        \advance\leftmargin\labelsep\usecounter{enumi}}}
 \relax
\def\reflist{\section*{References\markboth
        {REFLIST}{REFLIST}}\list
        {[\arabic{enumi}]\hfill}{\settowidth\labelwidth{[999]}
        \leftmargin\labelwidth
        \advance\leftmargin\labelsep\usecounter{enumi}}}
 \relax
\makeatletter
\newcounter{pubctr}
\def\publist{\@ifnextchar[{\@publist}{\@@publist}}
\def\@publist[#1]{\list
        {[\arabic{pubctr}]\hfill}{\settowidth\labelwidth{[999]}
        \leftmargin\labelwidth
        \advance\leftmargin\labelsep
        \@nmbrlisttrue\def\@listctr{pubctr}
        \setcounter{pubctr}{#1}\addtocounter{pubctr}{-1}}}
\def\@@publist{\list
        {[\arabic{pubctr}]\hfill}{\settowidth\labelwidth{[999]}
        \leftmargin\labelwidth
        \advance\leftmargin\labelsep
        \@nmbrlisttrue\def\@listctr{pubctr}}}
 \relax
\makeatother

\def\be{\begin{equation}}
\def\ee{\end{equation}}
\def\ba{\begin{eqnarray}}
\def\ea{\end{eqnarray}}

\def\del{\partial}

\def\bx{{{\bar x}_{\scalebox{0.6}{$T$}}}}
\def\bt{{\bar t}}
\def\xc{{x_{\scalebox{0.6}{$T$}}}}

\def\XXint#1#2#3{{\setbox0=\hbox{$#1{#2#3}{\int}$}
     \vcenter{\hbox{$#2#3$}}\kern-.5\wd0}}

\def\Xint#1{\mathchoice
   {\XXint\displaystyle\textstyle{#1}}%
   {\XXint\textstyle\scriptstyle{#1}}%
   {\XXint\scriptstyle\scriptscriptstyle{#1}}%
   {\XXint\scriptscriptstyle\scriptscriptstyle{#1}}%
   \!\int}

\def\r{\rho}

\def\b{\beta}

\def\m{\mu}

\def\l{\lambda}

\def\bk{{\bf k}}

\def\no{\noindent}

\def\qq{\qquad}

\def\IR{\relax{\rm I\kern-.18em R}}

\def\inv{^{\raise.0ex\hbox{${\scriptscriptstyle -}$}\kern-.05em 1}}

\def \ha {{\frac{1}{2}}}

\def \ov {\over}

\def\half{{\textstyle {1 \over 2}}}

\newcommand{\bb}{\hskip -0.1cm}

\def\red{\textcolor[rgb]{0.98,0.00,0.00}}
\def\blue{\textcolor[rgb]{0,0.00,0.98}}

\begin{document}


\renewcommand{\theequation}{\thesection.\arabic{equation}}
\csname @addtoreset\endcsname{equation}{section}

\begin{titlepage}
\begin{center}

\renewcommand*{\thefootnote}{\arabic{footnote}}

\phantom{xx}
\vskip 0.3in



 {\large {\bf Triple critical point and  emerging temperature scales \\
  in $SU(N)$ ferromagnetism at large $N$}}

\vskip 0.3in

{\bf Alexios P. Polychronakos$^{1,2}$}\hskip .15cm and \hskip .15cm
{\bf Konstantinos Sfetsos}$^{3}$

\vskip 0.14in

${}^1\!$ Physics Department, the City College of New York\\
160 Convent Avenue, New York, NY 10031, USA\\
{\footnotesize{\tt apolychronakos@ccny.cuny.edu}}\\
\vskip 0.3cm
${}^2\!$ The Graduate School and University Center, City University of New York\\
365 Fifth Avenue, New York, NY 10016, USA\\
{\footnotesize{\tt apolychronakos@gc.cuny.edu}}

\vskip .14in

${}^3\!$
Department of Nuclear and Particle Physics, \\
Faculty of Physics, National and Kapodistrian University of Athens, \\
Athens 15784, Greece\\
{\footnotesize{ ksfetsos@phys.uoa.gr}}\\

\vskip .3in

\vskip .2in

\end{center}


\centerline{\bf Abstract}

\vskip .2in

\no
The non-Abelian ferromagnet recently introduced by the authors, consisting of atoms in the fundamental
representation of $SU(N)$, is studied in the
limit where $N$ becomes large and scales as the square root of the number of atoms $n$. This model exhibits
additional phases, as well as two different temperature scales related by a factor $N\!/\!\ln N$. The
paramagnetic phase splits into a "dense" and a "dilute" phase, separated by a third-order
transition and leading to a triple critical point in the scale parameter $n/N^2$ and the temperature, while the
ferromagnetic phase exhibits additional structure, and a new paramagnetic-ferromagnetic metastable phase
appears at the larger temperature scale. These phases can coexist, becoming stable or metastable as
temperature varies. A generalized model in which the
number of $SU(N)$-equivalent states enters the partition function with a nontrivial weight,
relevant, e.g., when there is gauge invariance in the system, is also studied and shown to manifest similar phases,
the dense-dilute phase transition becoming second-order in the fully gauge invariant case.

\vfill

\end{titlepage}
\vfill
\eject



\def\baselinestretch{1.2}
\baselineskip 20 pt

\newcommand{\eqn}[1]{(\ref{#1})}

\tableofcontents


\section{Introduction}
\label{intro}

Magnetic systems with higher internal $SU(N)$ symmetry are enjoying a revival in physics.
Such systems have been considered in the context of ultracold atoms \cite{Ghu,Gor,Zha,Mag,Cap,Mukherjee:2024ffz} or of interacting atoms
on lattice cites \cite{KT,BSL,RoLa,YSMOF,TK,Totsuka,TK2}, and were also studied
in the presence of $SU(N)$ magnetic fields \cite{DY,YM,HM}.

We have recently constructed a model for ferromagnets with $SU(N)$ degrees of freedom  which  manifests an intricate and nontrivial phase
structure \cite{Phases}. At zero magnetic field the system has three critical temperatures (vs. only one for $SU(2)$),
with a crossover between metastable states. Spontaneous breaking of the global $SU(N)$ symmetry arises in the
$SU(N) \to SU(N-1) \times U(1)$ channel at zero external magnetic field and generalizes to other
channels in the presence of non-Abelian magnetic fields. Further, due to the presence of metastable states,
the $SU(N)$ system exhibits hysteresis phenomena both in the magnetic field and in the temperature.

This raises the obvious question of whether a modification of the model and/or a different dynamical regime of the model
manifest novel features and phases, and this is the topic of the present work. The modification we will consider
involves weighing the partition function by a general power of the number of states of each irreducible
component of the global $SU(N)$ symmetry, thus modifying the entropy of each such component (the original
model corresponds to the power being 1). The new dynamical regime we will consider is the one where the rank of
the group $N$ grows large. Nontrivial modifications arise when the scaling is $N \sim \sqrt n$, with $n$ the number of
atoms in the magnet (playing the role of the volume of the system at the thermodynamic limit),  
generalizing and putting in a more physical context the infinite-temperature case studied in  \cite{Polychronakos:2023qwz}. The square root
scaling implies that modestly large $N$ (much smaller than $n$) will give rise to new effects.

As we shall demonstrate, the above modified ferromagnet (for zero external magnetic field) develops new phases,
splitting the paramagnetic phase of the finite-$N$ ferromagnet into a "dilute" (more paramagnetic)
and a "dense" (less paramagnetic) phase, separated from each other by a third-order phase transition and from
the ferromagnetic phase by a zeroth-order transition (the free energy is discontinuous) and leading to a triple critical point in the plane of the
scale parameter $N^2/n$ and the temperature. Further, an additional high temperature scale emerges, related to
the lower one by a factor of order $N\!/\! \ln N$, in which the ferromagnetic phase acquires additional structure and a new
"mixed" paramagnetic-ferromagnetic phase appears. These phases can coexist over a
range of parameters, the ferromagnetic one dominating thermodynamically at lower temperatures, becoming
metastable, and eventually disappearing at significantly larger temperatures, while the mixed phase remains metastable
at all temperatures. These features persist for all weighting factors for $SU(N)$-equivalent states,
with the notable difference that the
dense-dilute transition becomes second-order in the "gauge invariant" limit in which the number of states of each
irreducible component drops from the partition function (its power becomes zero).

The organization of the paper is as follows: In section \ref{genmod} we present the essential features of the model
and summarize the group theory tools needed for its analysis \cite{Polychronakos:2023yhq,Phases}.
In section \ref{lumpdildisti} we analyze the dilute (most paramagnetic) phase and derive
its critical transition temperatures; in section \ref{lumpdendisti} we perform the analysis of the dense
(less paramagnetic) phase and derive its critical temperatures; and in section \ref{disjph} we analyze the
ferromagnetic phase. In section \ref{w1w2} we study two special cases of the weighting power for the states, the
regular one fully taking into account all states, as in standard ferromagnets, and the one with full gauge invariance
between states in the same irreducible component, and derive analytical expressions for their transition temperatures. 
In section \ref{thephtr} we perform the thermodynamic analysis of the various phases, identify the emerging
high temperature scale, uncover the "mixed" metastable paramagnetic-ferromagnetic phase, and determine the
order of phase transition between the various phases. Finally, in section \ref{concl} we present our conclusions.

\section{The general model}
\label{genmod}

We consider a set of $n$ atoms, each carrying the fundamental representation of $SU(N)$ and interacting with
ferromagnetic-type interactions. Denoting by $j_{r,a}$ the $N\times N$-dimensional generators of $SU(N)$ in
the fundamental representation acting on atom $r$ at position $\vec r$, the interaction Hamiltonian of the full system is
\be
H = \sum_{r,s =1}^n c_{{\vec r},{\vec s}} \sum_{a= 1}^{N^2 -1} j_{r,a} \, j_{s,a}\ ,
\ee
where $c_{{\vec r},{\vec s}} = c_{{\vec r},{\vec s}}$ is the strength of the interaction between atoms $r$ and $s$.
This Hamiltonian involves an isotropic quadratic coupling between the fundamental generators of
the $n$ commuting $SU(N)$ groups of the atoms. 
Assuming translation invariance $c_{{\vec r},{\vec s}} = c_{{\vec r}-{\vec s}}$, and also that the mean-field
approximation is valid,\footnote{The validity of the mean field approximation is
strongest in three dimensions, since every atom has a higher number of near neighbors and the statistical
fluctuations of their averaged coupling are weaker, but is expected to also hold in lower dimensions.}
each atom will interact with the average of the $SU(N)$ generators of the remaining atoms; that is,
\be
\sum_{{\vec r},{\vec s}} c_{{\vec r}-{\vec s}} \, j_{{\vec r},a} \, j_{{\vec s},a} = \sum_{\vec r} j_{{\vec r},a}
\sum_{\vec s} c_{\vec s}\, j_{{\vec r}+{\vec s},a}  \simeq \sum_r j_{r,a}
\Bigl(\sum_{{\vec s}} c_{\vec s} \Bigr)\,
{1\over n} \sum_{s'=1}^n j_{s',a} = -{c \over n} J_a \, J_a\ ,
\label{meanfield}\ee
where we defined the total $SU(N)$ generator
\be
J_a = \sum_{s=1}^n j_{s,a}
\ee
and the effective mean coupling
\be
c = - \sum_{{\vec s}} c_{\vec s}\ .
\ee
The minus sign is introduced such that ferromagnetic interactions, driving atom states to align,
correspond to positive $c$. Altogether, the effective interaction is proportional to the quadratic Casimir
of the total $S(N)$ generators
\be
H = -{c \over n} \sum_{a=1}^{N^2 -1} J_a^2\ .
\label{H0}\ee
Calculating the partition function involves decomposing the full Hilbert space of the tensor product of $n$
fundamentals of $SU(N)$ into irreducible representations (irreps), each of which has a fixed quadratic Casimir. This is
most conveniently done in the fermion momentum representation. Specifically, to each irrep with Young tableau row
lengths $\ell_1 \geqslant \ell_2 \geqslant \cdots \geqslant \ell_{N-1}$ we map a set of $N$ distinct non-negative
integers $k_1 > k_2 > \cdots > k_N$ such that
\be
\label{lengyt}
k_i - k_N = \ell_i +N - i\ ,\qq  {\sum_{i=1}^N k_i }= n+{N(N-1)\over 2}\ .
\ee
If $n < N$, $k_N = 0$, but for $n \geqslant N$ all $k_i$ may become nonzero.
We label each irrep with its corresponding vector $\bk = \{k_1,\dots,k_N\}$.
The relevant quantities for our ferromagnetic model are:\\
$\bullet$
The dimension (number of states) $\dim(\bk)$ of the irrep labelled by $\bk$, given by
\be
\label{vandd}
\dim(\bk) = {\Delta({\bf k}) \over \,\displaystyle{\prod_{s=1}^{N-1} s!}}~,\quad\text{with}~~~ \Delta(\bk) = \prod_{j>i=1}^N (k_i - k_j)\ .
\ee
$\bullet$ 
The quadratic Casimir $C_2 (\bk)$ of irrep $\bk$, given by
\be
C_2 (\bk) = {1\over 2}\sum_{i=1}^N k_i^2 -{1\over 2}\left({n\over N} +{N-1\over 2}\right)^2
-{N(N^2 -1)\over 24}\ .
\label{caso}
\ee
$\bullet$
The multiplicity $d(n;\bk)$ of the irrep $\bk$ in the decomposition of $n$ fundamentals, given by \cite{Polychronakos:2023yhq}
\be
\label{caso5} 
d(n;\bk) = {n!\, \Delta({\bf k}) \over \displaystyle\prod_{i=1}^N k_i !}\ .
\ee
With the help of the above, the partition function of the model in temperature $T = \beta^{-1}$, consisting of
the sum of the Boltzmann factors $e^{-\beta H}$ with $H$ as is \eqn{H0} over all $N^n$ states, can be written as an
explicit sum over $k_i$.

In what follows, we will consider a more general model in which each Boltzmann factor is weighted by a {\it power} of
the number of states $\dim(\bk)$ in each irrep. 
Omitting the irrelevant additive constants in the Casimir \eqn{caso}, our generalized partition function is given by
\be
Z_{w,n} = \sum_{\bf k} m(w,n;{\bf k}) \exp\bb\Big({\textstyle{{\b c\ov 2 n} \sum_{i=1}^N k_i^2}}\Big)\ , 
~\quad {\sum_{i=1}^N k_i }= n+{N(N-1)\over 2}\ ,
\label{Zmw}\ee
where $m(w,n;\bk)$ is the generalized thermodynamic multiplicity\footnote{
This factor was introduced in \cite{Polychronakos:2023qwz} where the infinite
temperature limit of our model was studied and shown to have a rich structure depending on the value of $w$.
A possible additional power in the multiplicity $d(n;\bk)$ can be absorbed into a redefinition of the remaining
parameters of the model in the thermodynamic limit.}
\be
\label{mww}
m(w,n;{\bf k}) =\dim(\bk)^{w-1} d(n;\bk) = 
{n!\, \Delta({\bf k})^w \over \,\displaystyle{\Big(\prod_{s=1}^{N-1} s! \Big)^{w-1} \prod_{i=1}^N k_i !}\,}\ .
\ee
The parameter $w$ assigns an exponential weight to the number of states of each irrep, $w=2$ corresponding to the standard
thermodynamic ensemble. Other values of $w$ are relevant in specific contexts. In particular, $w=1$ would correspond to a
situation where $SU(N)$ is a gauge group, and all states transforming in the same irrep are gauge-equivalent and count
as a single state.
The case $w<1$ is rather unphysical, since in that case the thermodynamic contribution of an irrep would {\it diminish} with its
number of states. In what follows we are mostly interested in the cases $w=1$ and $w=2$, but we keep the discussion
general.

\no
As explained in the introduction, when $N^2$ and $n$ are comparable, the phase structure of the model changes compared
to the standard thermodynamic limit $N^2 \ll n$, $w=2$. What happens in general is that the unbroken (paramagnetic)
phase of the standard ferromagnet maps to a phase with a distribution of irreps parametrized by a continuous density
of the $k_i$, while the broken (ferromagnetic) phase corresponds to a continuous distribution plus one isolated $k_i$, which turns out to be the largest one $k_1$.
We call the first one "lumped" and the second one "disjoint" phases. The lumped (continuous) distribution
can either saturate the density limit of 1 for a range of values of $k$ (corresponding to a set of adjacent $k_i$), or never saturate it. We call the first one "dense" and the second one "dilute" phases. 
At high temperatures, the disjoint distribution can either correspond to a single-row symmetric irrep, or also have a lumped part corresponding to nonzero lower Young tableau rows. We call
the first one simply disjoint, or symmetric, and the second one "lumped-disjoint", or "mixed" phases. Overall, we have three possible phases at lower temperatures 
(the dilute, the dense, and the disjoint phases)
and four possible phases at very high temperatures (the dilute, the dense, the disjoint, and the lumped-disjoint phases), which can coexist. 
Later in the paper we will analyze the various phases and derive their thermodynamic transitions.

\section{Lumped dilute distributions}
\label{lumpdildisti}

In the large-$n,N$ limit it is possible and convenient to describe the set of vectors $\bk$ in terms of the density $\r(k)$ of the
$k_i$ on the real positive axis. Writing the summand in \eqn{mww} as $e^{-\b F_{w,n}}$,
with $F$ representing the free energy of configuration $\bk$, and using the Stirling formula for factorials, we have (for details
on passing to the continuum we refer the reader to \cite{Polychronakos:2023qwz})
\be
\label{fun1}
\begin{split}
\beta F_{w,n}[\r(k)] = &  - {w \ov 2}  \int_0^\infty \bb\bb dk \bb \int_0^\infty \bb\bb dk' \, \rho(k) \rho(k') \ln |k - k' | 
\\
\quad  & +  \int_0^\infty \bb dk\, \rho(k) \Big(k (\ln k -1)- {\beta c\ov 2 n} k^2\Big) + (w-1) \sum_{s=1}^{N-1} \ln s! -\ln n!\ ,
\end{split}
\ee 
where the last two terms are constant and will be  omitted in the subsequent analysis.
We will consider distributions $\r(k)$ that are everywhere differentiable, except at points where $\rho(k)$ becomes $0$ or $1$.

\no
In the thermodynamic limit $n\gg1$, the sum (integral) over $k$ will be dominated by its largest term, that is,
the term minimizing $F_{w,n}$, under the constraints 
\be
\label{constr}
\int_0^\infty \bb dk\, \rho(k) = N  \ ,\qq 
 \int_0^\infty \bb dk\, k\, \rho(k) = n + {N^2\over 2}\ ,\qq 0\leqslant \r(k) \leqslant 1\ ,
\ee
which fix the rank of the group $SU(N)$ and the number of atoms $n$, and implement the fact that the
density of $k_i$ can neither be negative nor exceed one, since $k_i -k_{i+1} \geqslant 1$.
These constraints already suggest the scaling $N^2 \sim n$. We thus define the rescaled variable and parameters
\be
\label{torde}
x = {k\ov N} ~,\qquad  t = {4n\ov N^2} ~,\qquad T_0 = {c\ov N}
\ee
and keep $t,T_0$ finite as $n,N \gg1$ (the coefficient $4$ in $t$ is included to conform with the conventions
of \cite{Polychronakos:2023qwz}).
We also define a convenient temperature-dependent dimensionless parameter
\be
\label{lclc0}
\xc={tT\ov 4T_0}\ .
\ee
 In terms of the above, the constraints become 
\be
\label{constrt}
\int_0^\infty \bb dx\, \rho(x) = 1  \ ,\qq 
 \int_0^\infty \bb dx\, x\, \rho(x) = {t\ov 4} + {1\over 2} \ ,\qq 0\leqslant \r(x) \leqslant 1 \ .
\ee
Also, the functional \eqn{fun1} becomes 
\be
\label{fwnx}
\begin{split}
 \beta  F_{w,n}[\r(x)]   & =    N^2 \bigg[-{w \ov 2}   \int_0^\infty \bb\bb dx \bb \int_0^\infty \bb\bb dx' \, \rho(x) \rho(x') \ln |x - x' | 
\\
 &\ + \int_0^\infty \bb dx\, \rho(x) \Big( x (\ln x -1)- { x^2 \ov 2 \xc} \Big) 
 \bigg]\ ,
\end{split}
\ee
where we have neglected terms proportional to the constraints. 
We see that $F_{w,n}$ is indeed of order $N^2$,
justifying the saddle point approximation.
Enforcing the constraints \eqn{constrt} through similarly scaled Lagrange multipliers $\lambda,\mu$, we need
to extremize
\be
\beta F_{w,n}[\r(x)]  + N^2 \lambda \left( \int_0^\infty \bb dx\, x\, \rho(x) - {t\ov 4} - {1\over 2}\right)
+N^2 \mu \left(\int_0^\infty \bb dx\, \rho(x) - 1 \right)\ .
\ee
Variation in $\r(x)$ gives
\be
w \int_0^\infty dx'\, \rho(x') \ln|x-x'| = x(\ln x -1)  - {x^2\ov 2 \xc} +{\lambda}\,  x+ \mu\ .
\label{r2}
\ee
Further differentiating \eqn{r2} with respect to $x$ we obtain
\be
w\int_0^\infty dx' {\rho(x')\over x - x'} = \ln x - {x\ov x_ T} + {\lambda}\ .
\label{eomr}
\ee
The above equation must hold for $x$ such that  $0< \rho (x) < 1$.

\begin{figure} [th!]
\begin{center}
\includegraphics[height= 4.5 cm, angle=0]{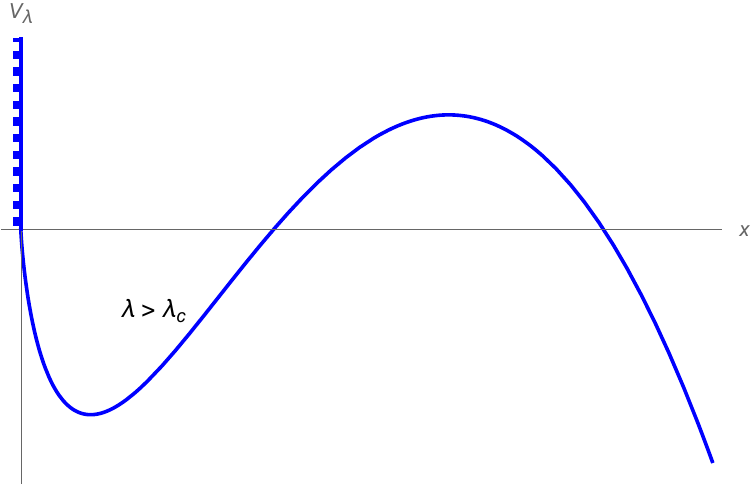}
\hskip 0.8 cm \includegraphics[height= 4.5 cm, angle=0]{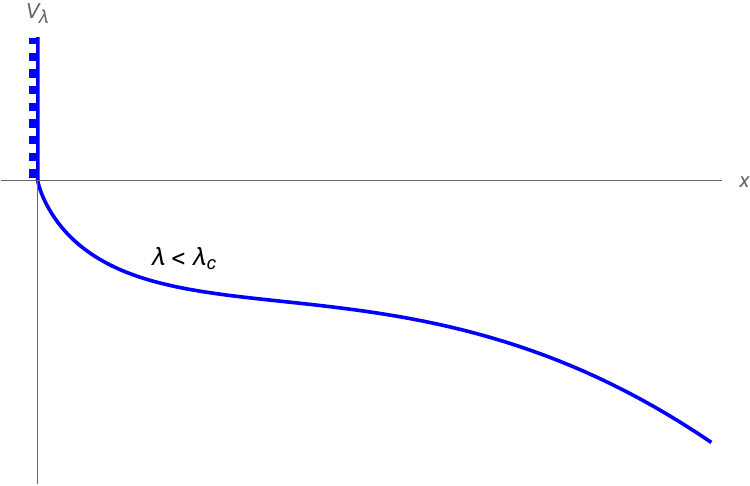}
\end{center}
\vskip -.5 cm
\caption{\small{The potential $V_\lambda (x)$ (shifted by the constant $\m$) for a generic value of $\lambda>\l_c$ (left) and $\l<\l_c$ (right). It has a "rigid wall" at $x=0$.}}
\label{PotentialVk}
\end{figure}

\begin{figure} [th!]
\vskip -0.5cm\begin{center}
\includegraphics[height= 5 cm, angle=0]{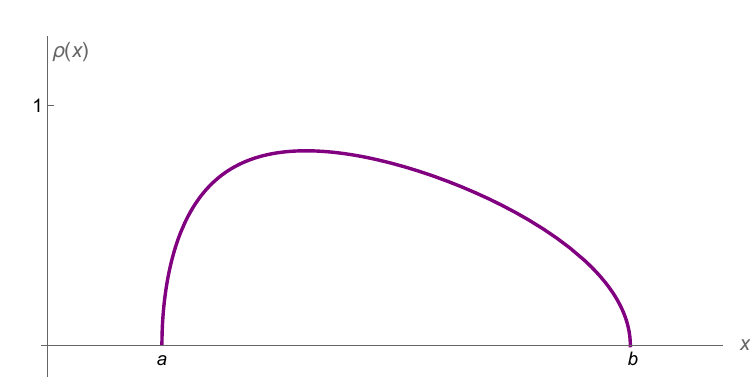}
\end{center}
\vskip -.5 cm
\caption{\small{A typical shape of the distribution $\rho(x)$ in the dilute phase.}}
\label{rhodilute}
\end{figure}

From (\ref{r2}) we see that the system corresponds to the equilibrium configuration of a
large number $N$ of particles with coordinates $x_i$ repelling each other with a logarithmic potential 
$w \ln|x_i - x_j|$ inside an external potential $V_\l (x)$ given by the right hand side of (\ref{r2}), that is 
\be
\label{potvl}
V_\l(x)=x(\ln x-1) - {x^2 \ov 2 \xc}+{ \lambda}\,  x +\mu\ .
\ee
Its derivatives
\be
V'_\l(x) = \ln x - {x\ov \xc} +{\lambda} ,\qq V''_\l(x) = {1\ov x} - {1\ov \xc}\ ,
\ee
indicate that, if $\l > \l_c$, where
\be\label{lclc}
\l_c = 1-\ln{\xc}\ ,
\ee
$V_\l(x)$ has a minimum at a value $x<\xc$ and a maximum at $x>\xc$ given by the solutions of
\be
\label{dhujh1}
x\, e^{\l}= e^{x/\xc}\ ,
\ee
while if $\l<\l_c$, $V_\l(x)$ is monotonically decreasing, as shown in fig. \ref{PotentialVk}. So, for the potential to
have a "well" able to retain particles, we need $\l > \l_c$.
%
%
%
The solution of \eqn{eomr} depends on whether the inequality constraint $\r(x) \leqslant 1$ is saturated
in a finite domain. The dilute phase correspond to the constraint not be saturated. This also means that the distribution
$\rho(x)$ does not reach the "wall" on the left at $x=0$, since the infinite potential there would force it to reach its
saturation value $\r(x)=1$. Therefore, $\r(x)$ is nonzero inside an interval $0<a<x<b$ and vanishes outside (see fig. \ref{rhodilute}).
Then solving (\ref{eomr})
becomes a standard single-cut Cauchy problem. We define the resolvent
\be
\label{ukr1}
u(z) = w \int dx\, {\rho(x) \over z-x}\ ,
\ee
with $z$ on the upper complex plane. Its real and imaginary parts on the real axis reproduce $\rho(x)$ and its
Hilbert transform
\be
u(x+i\epsilon) = w \Xint{-}  dx' {\rho(x') \over x-x'} -iw\pi  \rho(x)\ .
\label{ukr}
\ee
Therefore, a function that is analytic on the upper half plane and its real part on the real axis equals
$\ln x -x/\xc +\lambda$, as in \eqn{eomr}, will equal $u(z)$ up to an additive constant, and its imaginary part will fix
$\rho(x)$. In standard fashion, we write
\be
u (z) = {1\over 2\pi i} \sqrt{(z-a)(z-b)} \oint ds\, {\ln s - s/\xc +\lambda \over (s-z) \sqrt{(s-a)(s-b)}}\ ,
\label{rescut}
\ee
where the contour winds in the clockwise direction around the cut of the square root but does not
include the singularity at $z$ and the cut of the logarithm (see fig. \ref{contour1})
\begin{center}
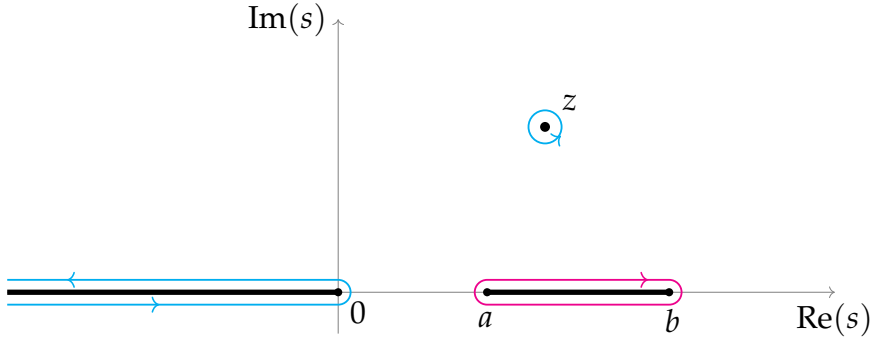

\begin{tikzpicture}[scale=1.1, decoration={markings,
mark=at position 0.9cm with {\arrow[line width=0.6pt]{<}},
mark=at position 7.4cm with {\arrow[line width=0.6pt]{<}}
}
]
\draw[help lines,->] (-4,0) -- (6,0) coordinate (xaxis);
\draw[help lines,->] (0,-0.5) -- (0,3.3) coordinate (yaxis);

\fill (2.5,2.0) circle (1.7pt);\fill (1.8,0) circle (1.4pt);\fill (4,0) circle (1.4pt);\fill (0,0) circle (1.4pt);

\node at (0.24,-0.25) {$0$};
\node at (2.8,2.3) {$z$};
\node at (1.78,-0.33) {$a$};
\node at (4.04,-0.34) {$b$};
\path[draw,line width=0.7pt,postaction=decorate] (2.3,2.0) [cyan] arc(180:-180:0.2);

\path[draw,line width=2.0pt] (1.8,0) 
-- (4,0)
;

\path[draw,line width=0.7pt,postaction=decorate] (4,-0.15) [magenta] arc(-90:90:0.15)
-- (1.8,0.15) arc(90:270:0.15) -- (4,-0.15) ;

\path[draw,line width=2.0pt] (-4,0) -- (0,0);
\path[draw,line width=0.7pt,postaction=decorate] (-4,0.15) [cyan] -- (0,0.15) arc(90:-90:0.15) -- (0,-0.15) -- (-4,-0.15);

\node[below] at (xaxis) {$\text{Re}(s)$};
\node[left] at (yaxis) {$\text{Im}(s)$};

\end{tikzpicture}
\vskip -.4 cm
\captionof{figure}{\small{Contour of integration in the $s$-plane. The original (magenta) contour
around the square root cut on $(a,b)$ is pulled back to the two (cyan) contours around the pole at $z$
and the logarithm cut on $(-\infty,0)$. In addition, there is the contribution of the circle at infinity, taken clockwise,
which is relevant for the linear in $s$ term in the numerator of the integrand \eqn{rescut}.}}
\label{contour1}
\end{center}
Pulling back the contour we pick up the pole at $s=z$ 
and the integrals around the cut of the logarithm and infinity
\be
\begin{split}
u (z) &= \ln z - {z\ov \xc} +{\lambda} 
\\
& \quad +{1\ov \xc} \sqrt{(z-a)(z-b)} -\sqrt{(z-a)(z-b)} \int_0^\infty {ds \over (s+z)\sqrt{(s+a)(s+b)}} \ .
\end{split}
\ee
Performing the integration we get that 
\be
\begin{split}
u(z) &= \ln z - {z \ov \xc} +{\lambda}\
\\
&\quad +  {i\ov \xc} \sqrt{(z-a)(b-z)}  - i \cos^{-1}{2z-a-b \over b-a} + i \cos^{-1}{(a+b)z - 2ab\over (b-a)z}\ .
\end{split}
\label{rescut1}
\ee
For $z =x$ real and between $a$ and $b$ (the region in which $\rho(x)$ does not vanish) 
the last three terms are purely imaginary (the square root factor multiplying the integral provides a
factor $+i$, since we assume that $z$ approaches $x$ from the upper-half complex plane). 
Then, according to (\ref{ukr}), we determine $\rho (x)$ as
\be
\begin{split}
\rho(x) &= {1\over w\pi}\Big( \cos^{-1}{2x-a-b \over b-a} -\cos^{-1}{(a+b)x - 2ab\over (b-a)x}\Big) - {\sqrt{(x-a)(b-x)}\ov w\pi \xc}
\\
&= {2\over w\pi} \cos^{-1} {\sqrt{x}+\sqrt{ab/x} \over \sqrt{a}+\sqrt{b}}  - {1\ov w\pi \xc} \sqrt{(x-a)(b-x)}\ ,\qq a\leqslant x\leqslant b\ .
\label{rhocos}
\end{split}
\ee
The second expression above makes clear that $\rho(x)$ vanishes at $x=a$ and $x=b$, and it never reaches or exceeds $1$ if $w\geqslant 1$ (as we assume). Plotting the density \eqn{rhocos} we indeed get a shape as in fig. \ref{rhodilute} above.
The parameters $a$, $b$ and $\lambda$ can be determined by matching the asymptotics of $u(z)$
\be
\label{assym}
\begin{split}
u(z) &=  wz^{-1} \int_0^\infty dx\, \rho(x) + wz^{-2} \int_0^\infty dx\, x\, \rho(x) + {\cal O}\big(z^{-3}\big)\, 
 \\
&= wz^{-1} + wz^{-2}\, \bigg({t\ov 4}+{1\ov 2}\bigg) +{\cal O}\big(z^{-3}\big)\ , 
\end{split}
\ee
where in the second line we used \eqn{constrt}. We obtain the conditions
\be
\label{Nnla}
\begin{split}
& 2w =  {(\sqrt{b}-\sqrt{a})^2}-{(a-b)^2\ov 4 \xc}\ ,
\\
&
w(t + 2) = {2(b-a)^2+(\sqrt{b}-\sqrt{a})^4\over 4} - {(a-b)(a^2-b^2)\ov 4 \xc}\ ,
\\
&{\lambda} = {a+b\ov 2 \xc}  -2 \ln{\sqrt{a}+\sqrt{b}\over 2} \ .
\end{split}
\ee
To proceed, it is convenient to define two new parameters as 
\be
p = {(\sqrt{a}+\sqrt{b})^2\ov 4 \xc} \ , \quad q= {(\sqrt{b}-\sqrt{a})^2\ov 4 \xc}  \ , \qquad p>q>0 \ ,
\label{pq}
\ee
with inverse 
\be
a= \xc (\sqrt{p}-\sqrt{q})^2\ ,\qq b= \xc (\sqrt{p}+\sqrt{q})^2\ .
\ee
Then, the system \eqn{Nnla} can be written more compactly as 
\be
\label{Nnla1}
\begin{split}
& 2 \bx\, q(1-p) = 1\ ,
\\
& 
{\bar t}+2 =  8 \bx^2\, q\big(p + {q \over 2} - pq -p^2 \big)  \equiv g(p,q) \ ,
\\
&{\l} = p+q - \ln(\xc p) \ .
\end{split}
\ee
where
\be
\bx = {\xc \over w}= {tT\over 4w T_0}~,\qquad {\bar t} = {t+2\over w} -2\ ,
\label{xtb}\ee
thus eliminating $w$ from the equations for $p,q$.
Note that the first equation in \eqn{Nnla1} gives the stronger restriction $0<q<p<1$ for $p$ and $q$.

\no
In order for the above picture to be valid we should have $\l>\l_c$, where $\l_c$ was defined in \eqn{lclc}. This implies that
\be
\label{cond0}
\quad  p-\ln p >1-q\ .
\ee
This  is identically satisfied if (\ref{Nnla1}) has real solutions. Indeed, since both $p$ and $q$ are positive, the left 
hand side of the inequality has a minimum of $1$ at $p=1$, while the right hand side is less than $1$.

\no
In addition, the density has to satisfy $0\leqslant \r(x)\leqslant 1$. Obviously, since the first term in \eqn{rhocos}
is bounded by unity and the second is negative, the density never exceeds unity. Nevertheless the density
can become negative for a range of values of $x\in [a,b]$ rendering the solution unacceptable.  
Demanding that the derivatives of $\r(x)$ at the end points at $x=a$ and $x=b$ are positive and negative,
respectively, and recalling that $b>a$,
we conclude that a necessary condition for having $\r(x)\geqslant 0$ in the interval $x\in [a,b]$, is
\be
\label{cond1}
\xc > \ha (b+\sqrt{ab}) \quad \Longrightarrow\quad p+\sqrt{p q} <1\ .
\ee
This condition is also sufficient for having $\r(x)\geqslant 0$. 
Indeed, setting $\r'(x)=0$ gives
\be
x^2 -\xc (1+p+q) x + \xc^2 (p-q) = 0\ ,
\label{maxomin}\ee
which has two real solutions for all values of the parameters in their range.  
Since $\r(x)$ starts up with positive (negative) derivative at $x=a$ ($x=b$), where it vanishes,
it cannot have two extrema between these values, and has just a single positive maximum.
The condition \eqn{cond1} ensures that the largest root of \eqn{maxomin}, which corresponds
to a negative minimum of $\r(x)$, is larger than $b$ and thus outside the range $(a,b)$.

\no
Solving for $q$ the first equation in (\ref{Nnla1}), the condition $p>q$ requires
\be
\label{pmpp}
p_- < p < p_+ \ ,\qq \bx >2\ ,
\ee
with
\be
p_\pm = {1\pm\sqrt{1-{2/ \bx}} \over 2}\ .
\ee
Then (\ref{cond1}) becomes
\be
f(p)\equiv p+\sqrt{p\over 2 \bx (1-p)}-1 <0\ .
\label{condom}
\ee
The function $f(p)$ is increasing, and we can see that
\be
f(p_\pm) =\pm \sqrt{1-{2\over \bx}} \ .
\ee
Therefore, (\ref{condom}) holds for 
\be
\label{pmpp0}
p_- <  p<p_0< p_+\ , \qq \bx >2\ \ ,\qq f(p_0 )=0\ ,
\ee
which is a refinement of \eqn{pmpp}.
The value of $p_0$ can be calculated analytically and arises from a cubic equation with one real solution given by
\be
\label{popo}
p_0 = 1-6^{-1/3} \zeta^{-1} +6^{-2/3} \bx^{-1} \zeta\ ,\qq \zeta 
= \Big(\sqrt{6 \bx^3 + 81 \bx^4}-9 \bx^2\Big)^{1/3}\ ,
\ee
and monotonically increases from $1/2$ to $1$ in the interval $\bx \in [2,\infty)$ and is always bounded by $p_+$.

\no
Finally,  the function $g(p) = g(p,q(p))$ defined in the second equation of (\ref{Nnla1}),
after eliminating $q$ using \eqn{Nnla1}, becomes
\be
g(p) = {1-2p \over (1-p)^2}+4\bx p \ .
\ee
Moreover, since
\be
g' (p ) =  4\bx - {2p\ov (1-p)^3}  \ ,
\ee
we easily see that $g'(p_0)=0$. Since $g''(p)<0$ the function $g(p)$ has a maximum at $p=p_0$.
It is easily checked that $g(p_-) >0$ and therefore $g(p)>0$ and increasing in $p\in [p_-,p_0]$. 
Consequently, the second of  \eqn{Nnla1} has a unique solution in $p$ for $\bar t$ in the range
\be
\label{gptp0}
g(p_-) < {\bar t} + 2 < g(p_0) 
\ee
and no solutions outside this range, signaling a transition to different phases.
For $g(p_-)$ we may use the expression
\be
g(p_-) =  2\Big(3\bx -\bx^2 +\sqrt{\bx (\bx-2 )^3}\Big)\ .
\ee
In addition, $g(p_0 )$ can be computed to be 
\be
\label{gp0}
g(p_0) = {4p_0^2 - 3p_0 +1\ov (1-p_0)^3}\ ,
\ee
where $p_0$ is given by \eqn{popo}.

The condition ${\bar t} + 2 > g(p_-)$ in \eqn{gptp0} derives from $p>q$, that is, $a >0$. Saturating it implies $a=0$ and signals
a transition to a dense phase, in which $\rho(x)$ extends to $x=0$ and starts saturating to $\rho(x)=1$. Therefore,
${\bar t}+ 2= g(p_-)$ identifies the critical temperature $T_c$ at which the transition happens.\footnote{Physically,  to be in the dilute phase we should have $T>T_c$. This can also be seen algebraically by expanding the left hand side of the inequality 
$g(p_-) - \bt -2< 0$ near $T=T_c$. To leading order this is proportional to 
$\displaystyle \Big(g'(p_-){dp_-\ov d \bx}\Big|_{T_c} + 4 p_-\Big) (T-T_c)<0$. The coefficient of $T-T_c$  is negative and therefore $T>T_c$ as stated.}
Working out the details of this equation we get a second order algebraic equation for $T$ whose  positive solution is the critical temperature 
\be
\begin{split}
\label{Tc}
& T_c =   2w T_0\, {3t +6 -8w + w^{-1/2}\big(4w-2 -t\big)^{3/2}\ov t(t-3w+2)}\ ,
\\
 &\qq\  3w-2\leqslant t\leqslant 4w-2\quad  (1\leqslant \bt \leqslant 2) \ ,
\end{split}
\ee
where we have also used \eqn{xtb}. The critical temperature $T_c$ is monotonically decreasing with $t$, from $T_c \to \infty$ as $t\to 3w-2$, 
consistent with the result of \cite{Phases}, to $T_c=4w T_0 /(2w-1)$ for $t=4w-2$.
If particular, near $t=4w-2$, we have the expansion 
\be
\label{tcn}
T_c = T_0 {4w\ov 2w-1}  - T_0  {4 w-1 \ov  (2 w-1)^2} (t - 4 w+2) + \dots\ .
\ee
and near $t=3w-2$ the behavior 
\be
T_c = T_0 {4 w^2 \ov  3 w-2}\, {1\ov t - 3 w+2} + \dots\ .
\ee
Finally, note that the corresponding critical value for $b$ is
\be
\label{bccr}
b_c = 4 w - 2 \sqrt{w} \sqrt{4w -2 - t}\ .
\ee
It increases with $t$, from $b_c =2 w $ as $t\to 3w-2$, to $b_c=4w$ for $t=4w-2$.

The condition $\bt + 2 <g(p_0)$ in \eqn{gptp0} derives from the positivity of the density $\rho$ and, when saturated, signals
an instability: the distribution $\rho (x)$ cannot be lumped and, instead, one isolated $k_i$ moves over
the hump of the potential to a large value of $k$. Therefore, $\bar t + 2= g(p_0)$ identifies another critical temperature
$T_s$ that marks the onset of instability and the transition to a disjoint phase.\footnote{Again for $T>T_s$ we are in the dilute phase. 
This can be seen also be seen by expanding the right hand side of the inequality 
$g(p_0) - \bt -2> 0$ near $T=T_s$. To leading order this is proportional to $\displaystyle \Big(g'(p_0){dp_0\ov d\bx}\Big|_{T_s} + 4 p_0\Big) (T-T_s)>0$. The coefficient of $T-T_s$  is positive and therefore $T>T_s$ as stated.}
Using \eqn{gp0} we have an explicit expression of $t$ in terms of $T_s$ which is
hard to invert. To proceed we recall that 
\be
 2 \bx = {p_0\over (1-p_0)^3 }
\label{ptb}
\ee
and reparametrize as
\be
p_0 = {s \over s+1}\quad \Longrightarrow \quad  2 \bx = s(s+1)^2\ .
\label{pxts}
\ee
Note that, \eqn{ptb} and the first of \eqn{Nnla1}  amount to saturating the constraint in \eqn{cond1}.
Proceeding, the condition $g(p_0)=\bt +2 $ upon using \eqn{gp0} and \eqn{pxts} gives
\be
s^2 (2s+1) = \bt +1\ ,
\ee
with primary solution
\be
s = {1\over 6}\left(\delta+\delta^{-1} -1\right) \ ,
\qq \delta = \left(\bb\sqrt{(54\bt+53)^2-1}+54\bt+53\right)^{1/3}\ .
\ee
Plugging in \eqn{pxts}, gives the critical temperature $T_s$ as
\be
T_s = T_0 {w\over 12t}\left(\delta^2 + 2\delta + 2 \delta^{-1} + \delta^{-2} +12\bt +11\right)\ ,\quad  t\geqslant 4w-2\quad  ( \bt \geqslant 2) \ .
\label{Ts21}
\ee
Note that, for $t=4w-2$, we have the expansion 
\be
\label{tsn}
T_s = T_0 {4w\ov 2w-1}  - T_0 {t + 2 - 4 w\ov  (2 w-1)^2} + \dots\ .
\ee
Therefore, $T_s$ joins $T_c$ at the highest $t$
for which $T_c$ is defined albeit with a first derivative discontinuity as it can be seen by the expansions \eqn{tcn} and \eqn{tsn}.
On the other hand, $T_s (t)$ for $t < 4w-2$ does not make sense, since it enters the region
$T<T_c$ in which the lumped solution does not hold.
For $t\to \infty$, $T_s$ slowly asymptotes to $T_0$ as
\be
T_s = T_0 + 3 T_0 \Big({w\ov 4 t} \Big)^{1/3} + \dots \ .
\ee

\section{Lumped dense distributions}
\label{lumpdendisti}

\begin{figure} [th!]
\begin{center}
\includegraphics[height= 6 cm, angle=0]{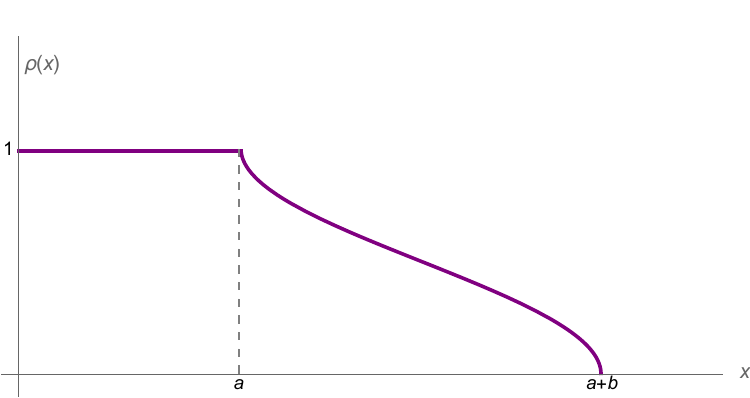}
\end{center}
\vskip -.5 cm
\caption{\small{A typical shape of the distribution $\rho(x)$ in the dense phase.}}
\label{rhodense}
\end{figure}

Below the critical temperature $T_c$, the distribution $\rho (x)$ touches $x=0$ and the dilute solution is not
valid any more. A finite part of the distribution will condense to the maximal value $\rho =1$ near $x=0$, and the
equilibrium conditions for the remaining $\rho<1$ part are modified.
Assuming that $\rho(x)$ condenses for $0<x<a$, and vanishes for $x>a+b$, we set
\be
\label{lupde}
\rho(x) = \left\{ \begin{array}{cc}
1\ , \qquad &0<x<a\ , \cr
\r_0 (x-a)\ , \qquad & a<x<a+b \ ,
\end{array} \right.
\ee
and zero elsewhere, with $a$ and $b$ positive constants (see fig. \ref{rhodense}). The constraints \eqn{constrt} 
become
\be
0\leqslant  \rho_0 (x) \leqslant 1 \ ,\quad \int_0^b dx\, \rho_0 (x) = 1-a\ ,\quad  
\int_0^b dx \, x\, \rho_0 (x) = {t\ov 4} + {(1-a)^2 \over 2}\ .
\label{cono}
\ee
Substitution of $\rho (x)$ in \eqn{fwnx} yields, upon changing variable $x \to x+a$,
\be
\begin{split}
\beta F_{w,n} [\rho_0 (x)]  = N^2 &\left[-{w\ov 2} \int_0^\infty \bb\bb dx \int_0^\infty \bb dx' \, \rho_0 (x) \rho_0 (x') \ln |x - x' | \right.
\\
& + \int_0^{\infty} \bb dx\, \rho_0(x)\, \Big(w x (\ln x -1)-{x^2\over 2\xc}\Big)   
\\
&\left. -(w-1) \int_0^{ \infty} dx\, \rho_0 (x)\, (x+a)  \big(\ln (x+a) -1\big)\right] \ ,
\end{split}
\ee
where we omitted terms that are set to constants by the constraints.
Implementing the constraints \eqn{cono} with appropriate Lagrange multipliers and minimizing the effective action leads to
the equilibrium equation
\be
\begin{split}
w \int_0^\infty\! dx'\, \rho_0 (x') \ln|x-x'| = &\, ~w x(\ln x -1)  -(w-1) (x+a)\big(\ln (x+a) -1\big)  
\cr
& - {x^2\ov 2 \xc} + {\lambda} x + \mu\ .
\label{r20}
\end{split}
\ee
Taking the $x$-derivative we obtain the analog of \eqn{eomr}, i.e.,
\be
w\int_0^\infty dx' {\rho_0 (x')\over x - x'} = w\ln x -(w-1) \ln(x+a) - {x\ov \xc}  + {\lambda} \ ,
\label{wo}
\ee
when $\rho_0 (x) > 0$. 
We see that now the equation for $\r_0(x)$ has a two-logarithm potential, given by the right hand side of
\eqn{r20} as
\be
V_{\l,w,a}(x) =  w\, x(\ln x -1)  -(w-1) (x+a)\Big(\ln (x+a) -1\Big)   -{x^2\ov 2 \xc} +{\lambda} x +\mu \ ,
\ee
while for $w=1$ the second logarithm drops out.

\no
To solve for $\r_0 (k)$, we define as before the resolvent
\be
u_0 (z) = w\int dk\, {\rho_0 (k) \over z-k}\ ,
\label{uoz}
\ee
reproducing $\rho_0 (k)$ and its Hilbert transform as
\be
u_0 (k+i\epsilon) = w \Xint{-} dk' {\rho_0 (k') \over k-k'} -iw\pi  \rho_0(k)\ .
\label{ukw}
\ee
In analogy to \eqn{rescut}, we set
\be
u_0 (z) = {1\over 2\pi i} \sqrt{z(z-b)} \oint ds\, {w\ln s - (w-1) \ln(s+a) 
+ {\lambda} -{s/ \xc } \over (s-z) \sqrt{s(s-b)}}\ ,
\label{resw}
\ee
where the contour winds in the clockwise direction around the cut of the square root $[0,b]$
but does not include the
singularity at $z$ nor the cuts of the logarithms (so it "threads" the real line at $k=0$, see fig. \ref{contour2}).
\begin{center}
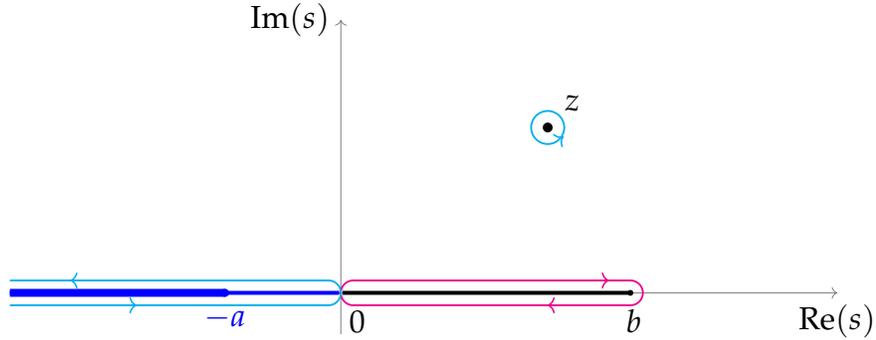

\begin{tikzpicture}[scale=1.1, decoration={markings,
mark=at position 0.9cm with {\arrow[line width=0.6pt]{<}},
mark=at position 7.4cm with {\arrow[line width=0.6pt]{<}}
}
]
\draw[help lines,->] (-4,0) -- (6,0) coordinate (xaxis);
\draw[help lines,->] (0,-0.5) -- (0,3.3) coordinate (yaxis);

\fill (2.5,2.0) circle (1.7pt);\fill (3.5,0) circle (1.0pt);\fill (-1.4,-0.0) [blue] circle (1.4pt);

\node at (0.2,-0.35) {$0$};
\node at (2.8,2.3) {$z$};
\node at (3.54,-0.34) {$b$};
\path[draw,line width=0.7pt,postaction=decorate] (2.3,2.0) [cyan] arc(180:-180:0.2);

\path[draw,line width=1.50pt] (0.02,0) 
-- (3.5,0)
;

\path[draw,line width=0.7pt,postaction=decorate] (3.5,-0.15) [magenta] arc(-90:90:0.15)
-- (0.15,0.15) arc(90:270:0.15) -- (3.5,-0.15) ;

\path[draw,line width=1.5pt] [blue] (-4,0.00) -- (-0.02,0);
\path[draw,line width=0.7pt,postaction=decorate] (-4,0.15) [cyan] -- (-0.15,0.15) arc(90:-90:0.15) -- (-0.15,-0.15) -- (-4,-0.15);

\path[draw,line width=3pt] (-4,-0.0) [blue] -- (-1.4,-0.0) node[below] {$-a$};

\node[below] at (xaxis) {$\text{Re}(s)$};
\node[left] at (yaxis) {$\text{Im}(s)$};

\end{tikzpicture}
\vskip -.4 cm
\captionof{figure}{\small{Contour of integration in the $s$ plane. The original (magenta) contour
around the (black) square root cut on $(0,b)$ is pulled back to the two (cyan) contours around the pole at $z$
and the two (blue) logarithm cuts on $(-\infty,-a)$ and $(-a,0)$.}
\label{contour2}}
\end{center}
Pulling back the contour we pick up the pole at $s=z$
and the integrals around the cuts of the logarithms and around infinity, and obtain
\be
\begin{split}
u_0  (z)  =&~ w\ln z -(w-1) \ln (z+a) -{z\ov \xc}+{\lambda} +{1\ov \xc} \sqrt{z(z-b)} \\
&-\sqrt{z(z-b)}\left[w\int_0^a +\int_a^\infty \right] {ds \over (s+z)\sqrt{s(s+b)}}\ .
\end{split}
\label{resw1}
\ee
For $z =x$ real and between $0$ and $b$ (the region in which $\rho_0(x)$ does not vanish) we obtain for
$\rho_0(x)$
\be
\begin{split}
\r_0(x) & = \sqrt{x(b-x)}\left[\int_0^a +{1\ov w} \int_a^\infty \right] {ds \over (s+x)\sqrt{s(s+b)}} 
 -{1\ov \pi w \xc} \sqrt{x(b-x)} 
\end{split}
\ee
and upon performing the integrals,
\be
\label{r0}
\r_0 (x) = {2\over w\pi} \cos^{-1} \sqrt{x \over b} + {2(w-1)\over w \pi}
\cos^{-1}\sqrt{(a+b) x \over (a+x)b} -{1\ov \pi w \xc} \sqrt{x(b-x)} \ .
\ee
The density $\r_0(x)$ obeys $\rho_0 (0) =1$ and $\rho_0 (b) = 0$.
The parameters $a$, $b$ and $\lambda$ can be related to $N$ and $n$ by matching the asymptotic 
expansion of $u_0 (z)$ 
\be
\label{assym2}
\begin{split}
u_0(z) &=  wz^{-1} \int_0^\infty dx\, \rho_0(x) +w z^{-2} \int_0^\infty dx\, x\, \rho_0(x) + {\cal O}\big(z^{-3}\big)\, 
 \\
&=w z^{-1}\, (1-a) +w z^{-2}\, \bigg({t\ov 4}+{(1-a)^2\ov 2}\bigg) +{\cal O}\big(z^{-3}\big)\ , 
\end{split}
\ee
obtained from (\ref{uoz},\ref{cono}) to those obtained from (\ref{resw}). We get
\be
\label{laaa}
\begin{split}
& 1-a = {b\over 2} - {w-1 \over 2w} \left(\sqrt{a+b}-\sqrt a\right)^2 -{b^2\ov 8 w \xc}\ ,
 \\
& {t\ov 2} + {(1-a)^2}  ={3b^2\over 8} + {w-1 \over 2w} \left(-{3b^2\over 4}+2 a^2
+(b-2a)\sqrt{a(a+b)}\right) - {b^3\ov 8 w \xc}\ ,
\\
& \lambda =  2 (w-1)\ln{\sqrt a + \sqrt{a+b} \over 2} -w \ln {b\ov 4} + { b\ov 2  \xc}\ .
\end{split}
\ee
We note that, for $a=0$, the equations \eqn{laaa} in the dense phase and \eqn{Nnla} in the dilute phase
become identical, since $b$ and $\lambda$ are continuous at the transition between the two phases, which
happens at the critical temperature $T_c$ in \eqn{Tc}.

The second critical temperature $T_s$ signaling the onset of instability occurs when $\r_0 (x)$ in \eqn{r0}
develops a negative part. Its extrema, happening at $\r_0' (x) =0$, satisfy
\be
\label{x2pm}
x^2+\Big(a-{b\ov 2}-\xc \Big) x -a\Big({b\ov 2}+\xc\Big) - (w-1) \xc \sqrt{a(a+b)} = 0 \ .
\ee
The smaller root $x_-$ is always negative and outside of the range of $x$ $[0,b]$. The positive root $x_+$ can be inside
the range and corresponds to a negative minimum. Stability is ensured if $x_+ >b$, giving the condition
\be
\label{2xct}
2 \xc > {b\sqrt{a+b}\ov \sqrt{a+b} + (w-1)\sqrt{a}}\ .
\ee
The critical point is at $x_+ = b$, which means that the above inequality is saturated, thus allowing to express the parameter $a$  as
\be
{\rm at\ criticality}: \qq a = {b(b-2\xc)^2 \over (2w \xc -b)\big(2(w-2) \xc +b\big)} \ ,\qq w\neq 1\ . 
\label{as}
\ee
Substituting the value \eqn{as} for $a$ in the first two equations in \eqn{laaa} determines $b$ and $T_{s'}$.
The equations are of high order and, in general, can only be solved numerically. It turns out that $T_{s'}$ joins $T_c$ and
$T_s$ for the dilute case at $t=4w-2$, forming a triple critical point on the $(t,T)$-plane, located at
\be
\label{trpoi}
{\rm Triple\ point}:\qquad t = 4 w -2 \, ,\quad T =   {4w\ov 2w-1}\,  T_0\ .
\ee
Moreover, it turns out that $T_{s}$ and $T_{s'}$ have their first 
derivatives (but not necessarily higher ones) continuous at $t=4 w-2$ and therefore join smoothly at this point. Hence, we may think of them as a single curve over the entire range of values of $t$, which we will denote by $T_s$, having 
different functional forms on either side of $t=4 w-2$ .

\section{The disjoint phase}
\label{disjph}

When $T < T_s$, either in the dilute or the dense lumped phase, the distribution develops a negative part,
signaling an instability. The stable configuration consists of
a single $k_i$ (the largest one) leaving the distribution and moving to a large value, as we shall see. 
Since the $k_i$ are ordered, we take this to be $k_1$. The remaining $k_i$ are treated using a distribution $\r(k)$
as before.

\begin{figure} [th!]
\vskip -0.5cm\begin{center}
\includegraphics[height= 5.4 cm, angle=0]{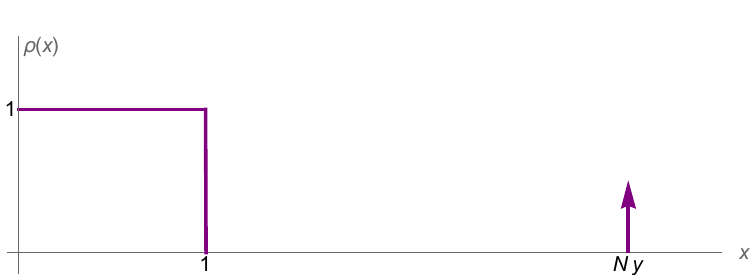}
\end{center}
\vskip -.5 cm
\caption{\small{The distribution $\rho(x)$ in the disjoint phase with one isolated $x_1 = N y$.}}
\label{rhodisjoint}
\end{figure}

It is clear that the contribution of the single isolated $k_1$ to the equilibrium equation for the remaining $k_i$ and
their density $\rho(k)$ is subleading in $N$. The only way that it can influence the distribution is through the
constraints \eqn{constr}, which become
\be
\label{constrr1}
\int_0^\infty \bb dk\, \rho(k) = N-1 \simeq N  \ ,\qq 
 \int_0^\infty \bb dk\, k\, \rho(k) = n + {N^2\over 2} - k_1\ .
\ee
This implies that $k_1$ must be of order $N^2$, and we set
\be
k_1 = N^2 y\ ,
\ee
contrasted to $k_i = N x_i$ for the rest of the $k_i$. 
The equilibrium equations for $\rho(x)$ and $y$ become
\be
\label{eomr1}
{w \over Nx-N^2 y}  + w\int_0^\infty dx' {\rho(x')\over x - x'}  = \ln x - {x\ov \xc} + {\lambda}\ 
\ee
and 
\be
\label{eomr2}
w\int_0^\infty dx { \rho(x) \over N y - x} = \ln (N y) - {N y  \over  \xc} +{\lambda} \ ,
\ee
where we also displayed the subleading terms. In addition, the constraints for $\r(x)$ become
\be
\int_0^\infty dx\, \rho(x) =1\ , \qq \int_0^\infty dx\, x\, \rho(x)= {t\over 4} + {1\over 2} - y\ .
\label{constrt1}
\ee
From \eqn{eomr1} we see that, indeed, $y$ does not influence the equation for $\rho$, as the $y$-dependent term
is of order $1/N^2$. Keeping leading terms in $N$, \eqn{eomr2} implies
\be
\lambda = N {y\over \xc}\ ,
\ee
so  that $\lambda$ assumes a macroscopically large (of order $N$) positive value. This drives the configuration for $\rho$
deep into the dense phase, that is,\footnote{That \eqn{ground} and \eqn{eomr33} are the solution to \eqn{eomr1}-\eqn{constrt} to leading order in $N$ also arises from the system \eqn{laaa} in the lumped dense phase upon shifting
$t/2\to t/2-2 y$, which takes into account the 
presence of the separate eigenvalue $k_1$ that modifies \eqn{constrr1} as compared to \eqn{constr}.
For large $\l$ the last of \eqn{laaa} implies that $b\to 0$. Then, the first two equations give 
$(1 - a)^2 + t/2 -y/2  + {\cal O}(b^2)=0$ and $1 - a +  {\cal O}(b)=0$, which implies that $a=1$ and $y=t/4$ as above
and the distribution \eqn{lupde} becomes the one of \eqn{ground}.} (see fig. \ref{rhodisjoint})
\be
\rho (x) =  \begin{cases}
1\,, \quad 0\leqslant x\leqslant 1\, ,
\\
0\, ,\quad {\rm otherwise}\ ,
\end{cases}
\label{ground}
\ee
and the second constraint in \eqn{constrt1} gives
\be
\label{eomr33}
 y={t\over 4}\ .
\ee
Therefore, the disjoint phase consists of a fully condensed distribution $\rho(x)$ and a single large
isolated $k_1 = N x_1 = N^2 t/4 = n$, the remaining $k_i$ ranging from $0$ to $N-2$.
The corresponding irrep is the fully symmetric one with a single row of
length $n$ in its Young tableau, so we will refer to the disjoint phase as the symmetric phase.

\no
We conclude by noting that, had we separated two or more large $k_i$, e.g., $k_1$ and $k_2$, we could not have
fulfilled the equilibrium condition. In the situations above, $k_1$ is sitting at the top of the effective potential
to leading order in $N$ (to subleading  order, slightly to the left, stabilized by the repulsion of the
lumped set $\rho(x)$). More than one $k_i$ would repel each other and could not stabilize. 
This is in accordance with the findings of \cite{Phases} where it was shown that, for finite $N$, configurations
with two or more rows in their Young tableaux lead to instabilities.

\no
The full phase diagram, obtained after deriving the corresponding phase transitions in section \ref{thephtr}, is
depicted in fig. \ref{PhaseDiag}.

\section{The special cases $w=1$ and $w=2$}
\label{w1w2}

As we mentioned in section \ref{genmod} on the general model, the cases $w=1$ and $w=2$ are special: $w=1$
corresponds to simply enumerating irreducible components, while $w=2$ is the standard
thermodynamic model where all states are counted with equal weight. As demonstrated in \cite{Polychronakos:2023qwz} these
values are also special from the mathematical point of view: in the absence of the energy term (when $T \to  \infty$),
they both admit simple, explicit solutions for the equilibrium configuration and its parameters $a,b$. Moreover,
the dense-dilute phase transition in the parameter $t$ is third order for $w=1$, and completely absent for $w=2$,
while for generic $w>1$ it is fourth order. We therefore examine the present model in these two special cases.

\subsection{$w=1$}

In this case the density in the dense phase \eqn{r0} specializes to 
\be
\label{ro1}
{\rm Dense}:\quad \r_0 (x) = {2\over \pi} \cos^{-1} \sqrt{x \over b} -{1\ov \pi \xc} \sqrt{x(b-x)} \ ,\qq 0\leqslant x \leqslant b\ ,
\ee
which obeys $\r_0(0)=1$ and $\r_0(b)=0$. 
This distribution remains in the physical range $0\leqslant \r_0(x)\leqslant 1$ for all values of $x\in [0,b]$
as long as
\be
\label{2xc1}
2\xc >  b \ ,
 \ee
 which follows by setting $w=1$ in \eqn{2xct}.
The equations determining the parameters $a,b$ and $\lambda$ \eqn{laaa} simplify to
\be
\label{laaa1}
\begin{split}
& 1-a = {b\over 2}  -{b^2\ov 8 \xc}\ ,
 \\
& {t\ov 4} + {(1-a)^2 \over 2}  ={3b^2\over 16} - {b^3\ov 16 \xc}\ ,
\\
& \lambda =  - \ln {b\ov 4} + { b\ov 2 \xc} \ .
\end{split}
\ee
This system can be solved for $a$ and $b$ by elimination, leading to a quadratic equation for $b^2$. The unique positive solution satisfying the
condition \eqn{2xc1} for the distribution $\r_0(x)$ to remain in the physical range, is given by 
$
b = 2\xc \sqrt{1-\sqrt{1-2t/\xc^2}}
$.
Then, $a$ and $\lambda$ are  obtained from the first and last equations in \eqn{laaa1}.
We are mostly interested in the expressions for the critical temperatures $T_c$ and $T_s$ which can be explicitly calculated.
The former  is obtained by setting $a=0$, which signals a transition to a dilute phase. From \eqn{laaa1} we obtain
\be
\label{gkok1}
T_c = 2T_0\, {3t -2 + (2-t)^{3/2}\ov t(t-1)}\ ,
\qquad 1\leqslant t\leqslant 2 \ ,
\ee
which of course is consistent with \eqn{Tc} for $w=1$ as it should be. 
The other critical temperature $T_s$ is obtained by setting $b$ to its critical value which, according to \eqn{2xc1} is at $b=2 \xc$.
The first equation in \eqn{laaa1} fixes $a=1-\xc/2$, and the second equation determines the critical temperature as
\be
T_{s }=  T_0  \sqrt{32\ov t} \ , \qq t\leqslant 2\ .
\label{Ts1}\ee
The range of $t$ is determined by demanding $a\geqslant 0$. For $t=2$, indeed $T_s =4 T_0$.
Also, $T>T_s$ warrants that the expression for $b$ given below \eqn{laaa1} remains real.

For the dilute phase, $\rho(x)$ is given by \eqn{rhocos} for $w=1$ which we reproduce for convenience here
\be
\begin{split}
{\rm Dilute}:\quad 
\rho(x) =  {2\over \pi} \cos^{-1} {\sqrt{x}+\sqrt{ab/x} \over \sqrt{a}+\sqrt{b}}  - {1\ov \pi \xc} \sqrt{(x-a)(b-x)}\ ,\quad a\leqslant x\leqslant b\ .
\label{rhocosd}
\end{split}
\ee
The constants $a,b,\lambda$ are determined by \eqn{Nnla} for $w=1$, with no substantial simplification. 
The critical temperature for transition to the disjoint phase is given
by \eqn{Ts21} for $w=1$ and ${\bar t} = t$, that is 
\be
\begin{split}
T_s = &  {T_0\over 12t}\left(\delta^2 + 2\delta + 2 \delta^{-1} + \delta^{-2} +12 t +11\right)\ ,
\\
& \quad   \delta = \left(\bb\sqrt{(54 t+53)^2-1}+54 t+53\right)^{1/3}\ ,\qq t\geqslant 2\ .
\label{Ts21g}
\end{split}
\ee
Its value and first derivative (but not the higher ones) match those of $T_{s}$ in the dense
phase \eqn{Ts1} above for $t=2$, the two curves essentially joining into one continuous curve for all values of $t$.

\subsection{$w=2$}

In this case a particular simplification and unification between the dense and dilute phases arises. The density in the dense phase \eqn{r0} specializes to 
\be
\label{ro2}
\r_0 (x) = {1\over \pi} \cos^{-1} {(1 + \sqrt{1+b/a})x - b
\over b\sqrt{1+x/a}}\  -{1\ov 2 \pi \xc} \sqrt{x(b-x)} \ ,\qq 0\leqslant x \leqslant b\ .
\ee
It obeys $\r_0(0)=1$ and $\r_0(b)=0$, and
remains in the physical range $0\leqslant \r_0(x)\leqslant 1$, for all values of $x\in [0,b]$ as long as 
\be
\label{jfijh2}
2 \xc >a+b-\sqrt{a(a+b)}\ ,
\ee
which follows by setting $w=2$ in \eqn{2xct}.
The system of equations \eqn{laaa} becomes
\be
\label{laaa2}
\begin{split}
& 4=\left(\sqrt{a+b}+\sqrt a\right)^2 -{b^2\ov 4 \xc}\ ,
 \\
& {t\ov 2} + {(1-a)^2}  ={3b^2\over 16} +{ a^2\ov 2} +{b-2a\ov 4}\sqrt{a(a+b)} - {b^3\ov 16 \xc}\ ,
\\
& \lambda =   { b\ov  2 \xc} -2 \ln{\sqrt{a+b} - \sqrt{a}\over 2}\ .
\end{split}
\ee
The density in the dilute phase \eqn{rhocos} specializes to
\be
\rho(x) = {1\over \pi} \cos^{-1} {\sqrt{x}+\sqrt{ab/x} \over \sqrt{a}+\sqrt{b}}  - {1\ov 2\pi \xc} \sqrt{(x-a)(b-x)}\ ,\qq a\leqslant x\leqslant b\ ,
\ee
which obeys $\r(a)=\r(b)=0$, and remains in the physical range $0\leqslant \r(x)\leqslant 1$ 
for all values of $x\in [a,b]$
as long as \eqn{cond1} holds. Repeated here for convenience this reads
\be
\label{jfij2}
2 \xc >b+\sqrt{ab}\ .
\ee
The system of equations \eqn{Nnla} becomes   
\be
\label{Nnla2}
\begin{split}
& 4 =  (\sqrt{b}-\sqrt{a})^2-{(b-a)^2\ov 4 \xc}\ ,
\\
&
2(t + 2) = {2(b-a)^2+(\sqrt{b}-\sqrt{a})^4\over 4} - {(a-b)(a^2-b^2)\ov 4 \xc}\ ,
\\
&{\lambda} = {a+b\ov 2 \xc}  -2 \ln{\sqrt{a}+\sqrt{b}\over 2} \ .
\end{split}
\ee
The critical temperature $T_c$ for transition between the two phases is given by \eqn{Tc} as
\be
T_c = 4 T_0\, {3t -10 + 2^{-1/2}\big(6 -t\big)^{3/2}\ov t(t-4)}\ ,\qq  4\leqslant t\leqslant 6
\ .\ee

The two systems \eqn{laaa2} and \eqn{Nnla2}), and the constraints \eqn{jfijh2} and \eqn{jfij2}, become, in fact,
identical upon using the $p,q$-parametrization:
\be
\begin{split}
&\text{Dense phase:}~~~a = \xc (\sqrt p - \sqrt q )^2 ~,~~ b = 4\xc \sqrt{pq} ~,~~ p<q\ ,
\\
&\text{Dilute phase:}~~~a = \xc (\sqrt p - \sqrt q )^2 ~,~~ b = \xc (\sqrt p + \sqrt q )^2~,~~ p>q\ .
\end{split}
\ee
Their common form is, upon combining the top two equations in \eqn{laaa2},
\be
\label{Nnpq}
\begin{split}
& \xc q (1-p) =1 \ ,
\\
&2\xc^2 \big(q^2 - 2pq (p+q-1)\big) = t+2\ ,
\\
&{\lambda} = p+q - \theta(q-p) (\sqrt p -\bb \sqrt q )^2 - \ln(\xc p)\ ,
\\
&p+\sqrt{pq} <1 \ ,
\end{split}
\ee
where $\theta$ is the Heaviside step function.
For $T>T_c$ the solution of \eqn{Nnpq} is for $p>q$, while for $T<T_c$ it is for $p<q$. So the transition
is completely analytic, the only sign of nonanalyticity being the Heaviside step function in the expression for the
Lagrange multiplier $\lambda$. This will be relevant in the study of the order of the transition.
The critical temperature $T_s$ for transition into the disjoint phase is given by \eqn{Ts21} 
for $w=2$, which are now valid for both the dense and dilute phases. Specifically, 
\be
\begin{split}
T_s =  &  {T_0\over 6t}\left(\delta^2 + 2\delta + 2 \delta^{-1} + \delta^{-2} +6 t -1\right)\ ,
\\
\quad  &  \delta = \left(\bb\sqrt{(27 t-1)^2-1}+27 t-1\right)^{1/3}\ ,
\qquad t \geqslant 0 \ .
\end{split}
\label{wTs21}
\ee
This diverges at $t=0$ as $T_s\simeq T_0 \sqrt{8/t}$ and slowly asymptotes to $T_s = T_0$ as $t\to \infty$.
At the triple point, $t=6$, $T_s = T_c = 8T_0/3$. 
The plots of $T_c,T_s$ are given in fig. \ref{PhaseDiag}.

\section{Thermodynamics and phase transitions}
\label{thephtr}

The free energy $F$, internal energy $U$, and entropy $S$ in each phase are related by the standard thermodynamic relations
\be
F= U-TS\ ,\qq U= -T^2 \del_T\bigg({F\ov T}\bigg)\ ,\qq S= - \del_T F\ .
\ee
The critical temperatures $T_c$ and $T_s$ define transition lines on the $(t,T)$ plane, with a triple point given in \eqn{trpoi}.
 To determine the order of these phase transitions, we examine the
thermodynamic quantities of each phase.

\subsection{Transitions between the dilute and dense phases}

For the dense and dilute cases, the expression \eqn{fwnx} for the free energy in terms of the distribution $\r(x)$
identifies the entropy and internal energy as
\be
\label{SU}
\begin{split}&S = N^2 \bigg[{w \ov 2} \int_0^\infty \bb\bb dx \bb \int_0^\infty \bb\bb dx' \, \rho(x) \rho(x') \ln |x - x' |  - \int_0^\infty \bb dx\, \rho(x) x (\ln x -1) 
\bigg]\ ,
\\
&U = -2 N^2{T_0\ov t} \int_0^\infty  dx\, x^2 \r(x) \ ,
\end{split}
\ee
calculated at the equilibrium configuration $\rho (x)$. Substituting the explicit expressions \eqn{rhocos} in the
dilute case, or \eqn{r0} in the dense phase, leads to some hard to evaluate integrals,
even after using the equilibrium equation \eqn{r2} to simplify $S$.
However, $U$ can be calculated from the asymptotic expansion of the resolvent
$u(z)$, as it is essentially the second moment of the distribution. This will allow us to determine the order of the
dense-dilute phase transition. Note that both $U$ and $S$, and consequently also $F$, are of order $N^2$.

For the lumped dilute phase, keeping one more term in the expansion \eqn{assym}, of order $z^{-3}$, we obtain
\be
\label{hfjh12}
\begin{split}
w \int_0^\infty  dx\, x^2 \r(x)  & =   {5(a^3+b^3) + 3 a b(a+b)  -2 \sqrt{a b}(3 a^2 +2 ab + 3 b^2)\ov 48}
 \\
 &  -{(a-b)^2 (5 a^2 + 6 ab + 5 b^2)\ov 128 \xc} \ .
\end{split}
\ee
Similarly, for the lumped dense phase we obtain 
 \be
w \int_0^\infty  dx\, x^2 \r(x) = {w\ov 3} a^3 + {5 b^3 \ov 48 } - {5 b^4\ov 128 x_T} 
+ {w-1\ov 24} {\sqrt{a}\, b^3(11 a + 3 b + 9 \sqrt{a(a+b)})\ov (\sqrt{a}+\sqrt{a+b})^3}\ ,
 \ee
where the first term comes from the unit part of the distribution in \eqn{lupde} and the rest from the nontrivial profile
$\r_0(x)$.

\no
We need to expand the above expressions in the temperature near the phase boundary at $T=T_c$, corresponding to $a=0$.
To do this, we expand $a$ and $b$ in $T-T_c$, which can be done by perturbatively solving the first two equations in \eqn{Nnla} and \eqn{laaa}. This expansion is qualitatively different for $w=1$ and $w>1$:
in the generic case $w>1$, the two sets of equations allow for an expansion of $\sqrt{a}$ and $b$ in powers of $T-T_c$
around their critical values $a=0$ and $b=b_c$. However, for $w=1$, the terms involving $\sqrt a$ in \eqn{laaa} drop,
and the variables to expand in the dense phase are $a$ and $b$. As we shall see, this also affects the order of the transition.
Hence, we treat separately the cases $w=1$ and $w>1$.

\subsubsection{$w=1$: $2^\text{nd}$ order phase transition }
 
First we consider the dilute phase and the system \eqn{Nnla} with $w=1$. It turns out that it suffices to keep terms of
${\cal O}(T-T_c)$ in the expansion of $\sqrt{a}$ and $b$. We find the expansion 
\be
\begin{split}
&  \sqrt{a}   ={1\ov  4 \sqrt{2} T_0}\, {t (1-  \sqrt{2-t})^2\ov (2-\sqrt{2-t})^{5/2}}\, (T- T_c) + \dots\ ,
\\
&
b =b_c +  {t(1-t)(1 - \sqrt{2 - t}) \ov 2 (2 - t (1 -\sqrt{2 - t}) ) T_0} \, (T- T_c) + \dots\ ,
\end{split}
\ee
from which the internal energy obtains as
\be 
\label{udil1}
U_{\rm dil} =U_0  + {N^2\ov 4} (1- \sqrt{2-t})^2(T-T_c) + \dots\ ,
\ee
where the internal energy at $T=T_c$ is
\be
\label{uow1}
U_0 = - {5 N^2\ov 12} T_c (t-1)\ .
\ee
This is negative, in accordance with \eqn{SU}, since approaching the critical curve $T=T_c$ requires that $1\leqslant t\leqslant 2$.

In the dense phase and the system \eqn{laaa1} we keep terms of ${\cal O}(T-T_c)$ in the expansion of
$a$ and $b$. We find that
  \be
\begin{split}
& a =  - {1\ov 2T_0}  {t (1-\sqrt{2-t})^2\ov   (2-\sqrt{2-t})(4-3\sqrt{2-t})}\, (T-T_c) + \dots\ ,
\\
&
b =b_c  + {(2-7 t +t^2)\sqrt{2-t} + t (2 + 5t -3 t^2)\ov (t^2-4)(2-9 t)T_0} \, (T-T_c) + \dots\ ,
\end{split}
\ee
from which
\be
\label{uden1}
U_{\rm den} =U_0 - {5 N^2\ov 4} {\sqrt{2-t}(1-\sqrt{2-t})^2\ov 4- 3 \sqrt{2-t}}\, (T-T_c)+ \dots\ .
\ee
Comparing \eqn{udil1} and \eqn{uden1} we see that there is a discontinuity in the first derivative of the internal energy with respect to $T$ at the critical temperature $T_c$. This corresponds to a second order phase transition.

\noindent
Note that the internal energy, as stated in \eqn{udil1} and \eqn{uden1}, is an increasing (decreasing) function of the temperature off the critical curve $T=T_c$ for the dilute (dense) phase.

  \subsubsection{$w>1$: $3^\text{rd}$ order phase transition}
 
In this case we will need to expand to order ${\cal O} (T-T_c )^2$. For the dilute phase, we find 
 \be
\begin{split}
& \sqrt{a} = { b_c^{3/2} T_0\ov 4 t T^2_c}\, (T-T_c) +
{b_c^{3/2} T_0 (2 b_c^3 T_0^3 - 7 b_c t^2 T_0 T_c^2 + 4 t^3 T_c^3)\ov 
 16 t^3 T_c^5 (2 b_c T_0 - t T_c)}\,  (T-T_c)^2  \dots\ ,
\\
&
b = b_c + {t(b_c-2 w)\ov 2 (2 b_c T_0 - t T_c)}\, (T-T_c) 
\\
&\qq   \qq -{b_c^2 T_0 (14 b_c^2 T_0^2 - 21 b_c t T_0 T_c + 
   8 t^2 T_c^2)\ov 16 T_c^2 (2 b_c T_0 - t T_c)^3}\, (T-T_c) ^2 + \dots \ ,
\end{split} 
\ee
with the critical values $T_c$ and $b_c$ given by \eqn{Tc} and \eqn{bccr}.
Then 
\be
\label{udilw}
\begin{split}
& U_{\rm dil} =  U_0 +N^2 {(b_c -2 w)^2\ov 16 w} \, (T-T_c) 
\\
& \qq \qq   -  N^2 {b_c (2 + t - 5 w) + 4 w^2\ov 8 (4 w-t-2) T_c}\, (T-T_c) ^2 + \dots\ , 
\end{split}
\ee
where the internal energy at $T=T_c$ is 
\be
\label{uoww}
U_ 0= -{5\ov 12} N^2 (t+2-3 w) T_c\ .
\ee
Note that setting $w=1$ in \eqn{uoww} recovers \eqn{uow1}, and that, again, $U_0<0$ in accordance with
\eqn{SU}.

For the dense phase the corresponding expansions are 
\be
\begin{split}
& \sqrt{a} =- {b^{3/2}_c T_0\ov 4 (w-1) t T_c^2}\, (T-T_c)  
\\
& \qq \
+ T_0 b_c^{3/2} {8 t T_c (w T_0-(w-1)^2 t T_c)+ T_0 b_c (8 w T_0 + (10-26 w + 13 w^2) t T_c)\ov 32 (w-1)^2 t^2 T_c^4 (2 T_0 b_c -t T_c)}
\,  (T-T_c)^2  \dots\, ,
\\
&
b = b_c + {t (b_c- 2 w)\ov 2 (2 b_c T_0 - t T_c)} \, (T-T_c)  - t(b_c-2 w) 
\\
& \times {8 (w-1)^2 t^2 T_c^2 +32 w^2 T_0^2 -4 w (8-14 w+7 w^2)t T_0 T_c - b_c T_0 ( 8 w T_0 
+ (6 -14 w + 7 w^2) t T_c)\ov 16 (w-1)^2 (2 b_c T_0 - t T_c)^3 T_c} \,
\\
& \qq\qq  \times (T-T_c) ^2 + \dots \ .
\end{split} 
\ee
\vskip -0.2in
\noindent
Then 
\be 
\label{udenw}
\begin{split}
&U_{\rm den} =   U_0 +N^2 {(b_c -2 w)^2\ov 16 w} \, (T-T_c)  -N^2 w (2 + t - 3 w) T_0 
\\
&
\times { (2 + t) (4 + 5 b_c + 2 t) - 
       4 w b_c (13 + 2 t)  + 4 w^2 ( b_c (18 + t) -8 (1 + t) )  + 
       16 w^3(2 - 2 b_c + t) \ov 8 t (b_c - 6 w) (2 + t - 4 w) (w-1)^2T_c^2} 
       \\
       & \qq\qq \times  (T-T_c)^2 \dots\ .
\end{split}
\ee
Comparing \eqn{udilw} and \eqn{udenw} we see that the first derivative of the internal energy is continuous at $T=T_c$
but there is a discontinuity in the second derivative. This corresponds to a third order phase transition.

\begin{figure} [h!]
\begin{center}
\includegraphics[height= 7.5 cm, angle=0]{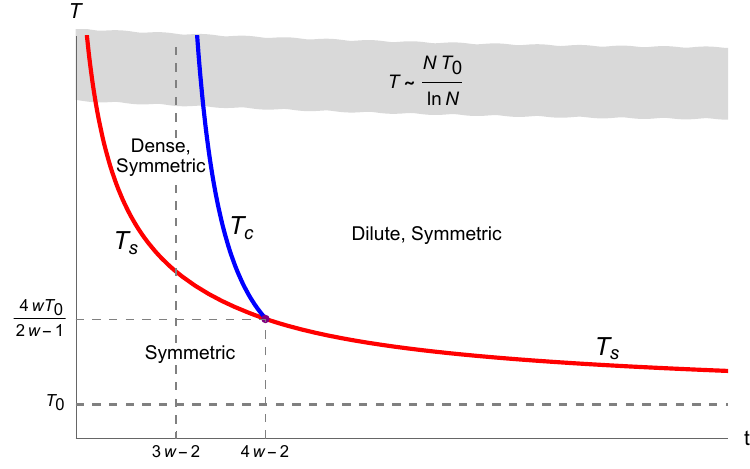} 
\end{center}
\vskip -.5 cm
\caption{\small{
Phase diagram in the $t \!-\! T$ plane with a triple point intersection given by \eqn{trpoi}. 
$T_s$ asymptotes to $T=T_0$ and $t=0$, while $T_c$ asymptotes to $t=3w-2$. 
The symmetric phase is everywhere stable, while the dense and dilute lumped phases are metastable for
$T>T_s$ and cease to exits for $T<T_s$. For high temperatures $T\sim T_0 N/\ln N$ as indicated in the gray area, there
is a nontrivial phase structure derived in section \ref{transjoin} and presented in fig. \ref{LargeTphases}.
}}
\label{PhaseDiag}
\end{figure}

\noindent
Also, unlike the $w=1$ case, the internal energy is an increasing function of the temperature off the critical curve $T=T_c$ for both the dilute and dense phases. In addition, it can be shown that  the sign of the $(T-T_0 )^2$ term in both expressions for the internal energy is negative. The phase diagram for temperatures $T \sim T_0$ is depicted in fig. \ref{PhaseDiag}.
\vskip -.2cm 

\subsubsection{An alternative computation}
The order of the dense-dilute transition can also be deduced from the Lagrange multiplier $\lambda$. Indeed,
the discontinuity of derivatives on the $(t,T)$ plane across the phase boundary $T=T_c$ is the same irrespective of
the direction in which we approach it. We can, thus, examine the derivative along the direction $tT\bb =\bb \text{constant}$,
which amounts to keeping the parameter $\xc = tT/4T_0$ in the free energy $F$ constant and differentiating with
respect to $t$. Since $F$ does not involve $t$ explicitly, we have
\be
\begin{split}
\left({\del F \over \del t}\right)_{\bb\xc} &= \int_0^\infty dx {\delta F[\rho(x)] \over \delta\rho(x)}\,
{\partial \rho (x) \over \del t} \\
\text{(using stationarity)} ~~~ &= -N^2 T \left(\lambda \int_0^\infty dx\, x {\partial\rho(x)\over \del t}
+ \mu \int_0^\infty dx\, {\partial \rho(x) \over \del t}\right) \\
& = -N^2 T \left(\lambda {\partial\over \del t}\int_0^\infty dx\, x \rho(x)
+ \mu {\partial\over \del t}\int_0^\infty dx\, \rho(x) \right) \\
\text{(using the constraints)}~~ &= -{N^2 T\over 4} \lambda\ .
\end{split}
\ee
From \eqn{Nnla} in the dilute phase and \eqn{laaa} in the dense phase we see that $\lambda_{\rm dil} = \lambda_{\rm den}$
at the phase boundary $a=0$.
For $w> 1$ the first derivatives $(\del \lambda_{\rm dil} /\del t)_\xc$ and
$(\del \lambda_{\rm den} /\del t)_\xc$ at $a=0$ are also the same. This can be checked explicitly for any $w\neq 1$,
 but it is easily seen from the form of $\l$ for $w=2$ in \eqn{Nnpq}, in which the continuity of the first derivatives
arises from the fact that the step function is multiplied by a term that vanishes quadratically as $p\to q$. Using similar reasoning, the second derivatives
$(\del^2 \lambda_{\rm dil} /\del t^2)_\xc$ and $(\del^2 \lambda_{\rm den} /\del t^2)_\xc$ differ at $a=0$, leading to
a discontinuity in $(\del^3 F /\del t^3)_\xc$ on the phase boundary and indicating a third-order phase transition.
For $w=1$ the absence of the first term in the expression for $\l$ in \eqn{laaa} leads to a discontinuity already in the
first derivative $(\del \lambda /\del t)_\xc$ on the phase boundary and thus to a discontinuity of
$(\del^2 F /\del t^2)_\xc$, indicating a second-order phase transition.

\subsection{Transitions to the disjoint phase and large temperatures}\label{transjoin}

For the disjoint phase, the free energy $F$ and energy $U$ can be calculated from the ground state distribution
\eqn{ground} and the isolated $k_1 = N^2 y = N^2 t/4 =n$. We see that the contribution of the isolated $k_1$
dominates, giving a free energy macroscopically larger (of order $N^3$) and an entropy macroscopically smaller
(of order $N\ln N$) than the ones in the other phases, namely\footnote{{These expressions follow from the exact
formulae of section \ref{genmod}.
For the fully symmetric irrep with $k_1=n+N-1$ and $k_i=N-i$, $i=2,3,\dots, N$, \eqn{vandd} gives
$\displaystyle \dim (\bk) ={(n+N-1)!\ov n!(N-1)!}\sim n^N N^{-N}\sim N^{ N}$, and \eqn{caso5} gives
$d(n,{\bf k})=1$,
as it should since there is only one way to compose $n$ fundamentals into a single row. 
Then \eqn{Zmw} yields $U$ and the logarithm of \eqn{mww} yields $S$.} \label{fff}}
\be
F_\text{disj} \simeq U_\text{disj} \simeq -{T_0 N\over 2n} n^2 = -{t\over 8}N^3 T_0\ ,\qq
S_\text{disj} \simeq (w-1) N\ln N\ .
\label{FSdisj}
\ee
Since the free energy of the lumped phases is of order $N^2$, the transition at $T=T_s$ involves a jump in the free energy and is of zeroth order. This also means that, at temperatures of order $T_0$,
the other two phases are metastable, since the disjoint phase has a macroscopically lower free energy.

The metastability frontier at which the free energy of the disjoint and either the dense or the dilute phase are equal would be at a
macroscopically large temperature $T \sim T_0 N/\ln N$. As we shall see, at such temperatures a new phase
appears, and the phase diagram in fig. \ref{PhaseDiag} acquires extra structure.

Assume a temperature of  the above order, that is,
\be
\label{Ttau}
T = \tau \, {N\over \ln N}  T_0\gg T_0 \ ,
\ee
with $\tau$ a dimensionless parameter of order 1. Substituting this value in \eqn{eomr2} and keeping leading-order
contributions leads to
\be
\lambda = \left( {4y\over t \tau} -1\right) \ln N + {\cal O}( 1 )\ .
\label{llnN}\ee
We also note that the second constraint in \eqn{constrt1} implies
\be
0<y \leqslant {t\over 4}\ ,
\ee
since the integral of $x \rho(x)$ is at least $1/2$.\footnote{A proof of this statement goes as follows: 
From the fact that $0<\r(x)<1$, we deduce that 
\be
\begin{split}
x > \int_0^x dy\, \r(y)\ & \Rightarrow\   \int_0^\infty dx\, x\r(x) >   \int_0^\infty dx\, \r(x)  \int_0^x dy\, \r(y) \\
 & = \ha  \int_0^\infty  d\Big[\int_0^x dy\, \r(y)\Big]^2 = \ha  \Big[\int_0^\infty dy\, \r(y)\Big]^2=\ha\ ,
\end{split}
\ee
where we used the first constraint in \eqn{constrt}. This bound arises from the corresponding discrete sum over $k_i$,
where the singlet representation with $k_i=0,1,\dots,N\bb-\bb1$ clearly provides the minimum value for the sum
$\sum_{i=1}^N k_i = N(N-1)/2\simeq N^2/2$, leading to the above bound in the continuum limit.
}

The value of the Lagrange multiplier $\lambda$, and the properties of the configuration, depend on the prefactor of
$\ln N$ in \eqn{llnN}. Specifically:

\begin{itemize}
\item 
\underline{$\tau > 4y/t$}: Then $\lambda$ is large ($\sim \ln N$) and negative. This destabilizes the equation for the
lumped part $\rho (x)$ and leads to no solution. Since $y\leqslant t/4$, we conclude that for $\tau >1$ there is no
solution and the disjoint phase disappears.

\item 
\underline{$\tau < 4y/t$}: Then $\lambda$ is large ($\sim \ln N$) and positive. This drives the configuration for the lumped
part $\rho(x)$ deep into the dense phase, that is, to \eqn{ground}, forcing $y$ to the value $y=1/4$. Therefore,
for $\tau <1$ the disjoint phase exists as discussed in detail in section \ref{disjph}.

\item 
\underline{$\tau = 4y/t$}: Then $\lambda$ becomes of order 1. Therefore, the equation for the lumped part $\rho(x)$
can have a nontrivial solution. In this case,
\be
\int_0^\infty dx\, x\, \rho(x)= {t\over 4} + {1\over 2} - y
= {t(1-\tau) \over 4} + {1\over 2} = {{\tilde t}\over 4} + {1\over 2}\ ,\qq \tilde t= t(1-\tau)\ .
\ee
Then, the equation for $\rho(x)$, becomes the standard one but with a high temperature as in \eqn{Ttau} and an effective ${\tilde t} = (1-\tau) t$, with the value of $\lambda$, which is now of order 1, adjusting
to the one reproducing the effective $\tilde t$.\footnote{The ${\cal O}(1)$ part of $\lambda$ is not fixed by \eqn{eomr2}
since it can be modified by adding subleading terms in the temperature, or equivalently changing $y$ by a
subleading term.}
The solution for the lumped part $\rho(x)$ is given by the
large-temperature limit of the solutions found in sections 3 or 4 (depending on the value of $\tilde t$), 
and it is also calculated in \cite{Polychronakos:2023qwz}, which examined the infinite-temperature limit of
the present model. This constitutes a new phase, which exists for all $0<\tau <1$, consisting
of a large 
$k_1 = N^2 \tilde t/4 = (1-\tau ) n$
plus a nontrivial lumped part $\rho$.

\end{itemize}

\no
Therefore, in the range $0<\tau<1$ there exist, a priori, three distinct phases: a lumped one ($y \simeq 0$),
corresponding to an $SU(N)$ irrep with a distribution of row lengths of order $N$ in its Young tableau;
a disjoint-lumped one ($y = t \tau/4$), corresponding to an irrep with one long row of order $n \sim N^2$
and a distribution of lower rows of order $N$; and a disjoint one ($y = t/4$), corresponding to a symmetric irrep
with a single row of length $n$. The first (lumped) phase is of paramagnetic nature, the second (mixed) one is
partially ferromagnetic, and the last (symmetric) one is fully ferromagnetic.

To determine the global stability and metastability properties of the above phases we need to compare their free energy:
\begin{itemize}

\item
The free energy of the disjoint phase is already calculated in \eqn{FSdisj}, which we reproduce here for convenience
\be
\label{Fdisj}
F_\text{disj} \simeq -{t\over 8} N^3 T_0\ .
\ee
\item
The free energy of the lumped phase at high temperatures $T \gg T_0$ is dominated by the entropy, that is,
$F_\text{lump} \simeq - T S$. Further, at high temperatures the entropy
approaches a limiting value $S_{\infty}$ that can be obtained from the results in this paper, but is also
already calculated in \cite{Polychronakos:2023qwz}. Dropping subleading terms in
(3.86) and (3.87) of \cite{Polychronakos:2023qwz} we obtain, for both the dense and dilute lumped phases,
$
S_{\infty} = n \ln N = {t\over 4} N^2 \ln N 
$,
which is the logarithm of the total number of states $N^n$ of the model for all values of $w$. 
Notably, $S_\infty$ is $w$-independent.
Therefore, using \eqn{Ttau} we obtain
\be
\label{Flump}
F_\text{lump} \simeq -\tau {t \over 4} N^3 T_0\ .
\ee

\item
The free energy of the lumped-disjoint phase can be calculated in a similar way. The end result consists of the
free energy due to the entropy of the lumped part $-T {\tilde n} \ln N$, with ${\tilde n} = n (1-\tau)$, plus the
energy of the isolated $k_1 = N^2 y = \tau n$, $U= - T_0 N k_1^2/(2n) = -\tau^2  t T_0 N^3/8$. Altogether,
\be
\label{Flump-disj}
F_\text{lump-disj} \simeq -\tau (2-\tau) {t \over 8} N^3 T_0\ .
\ee
\end{itemize}
\no
Comparing the three expressions in \eqn{Fdisj}-\eqn{Flump-disj} we see that the lumped-disjoint phase is always metastable. 
In addition, the disjoint phase is globally stable and the lumped phase is metastable for $\tau < 1/2$, while 
their roles are reversed
for $1/2<\tau <1$. For $\tau >1$ only the lumped phase exists.
As evident from \eqn{Tc} and fig. \ref{PhaseDiag}  (or from the results of  \cite{Polychronakos:2023qwz}, relevant in the present high-temperature regime), the lumped phase is in the dilute configuration for $t>3w-2$ and the dense one
for $t<3 w-2$.

The above picture admits a refinement. Indeed, the precise form of the disjoint phase changes slightly at $\tau = 1/(w+1)$.
Comparing the free energy of the fully symmetric irrep of length $n$, as in footnote \ref{fff}, to that of a configuration
with a number of boxes $s ={\cal O}(1)$ in its lower rows and a top row of length $n-s$, a direct computation
using the exact formulae of section \ref{genmod} yields a leading order change of free energy
\be
\Delta F \equiv F_{s\text{disj}}-F_\text{disj} \simeq s\Bigl(T_0 N - (w+1) T \ln N \Bigr)
= s N \Bigl(T_0 - (w+1) \tau \Bigr) \ ,
\ee
which is ${\cal O}(N)$, and expressed in terms of the relevant variable $\Delta x = s/N$ it is ${\cal O}(N^2)$,
while the free energy itself is ${\cal O}(N^3)$. Therefore, $\Delta F \simeq 0$ to leading order in $N$, as expected
from the fact that the disjoint is a local extremum.  Nevertheless, this indicates that for
\be
\tau > {1\over w+1}\qq  \Longleftrightarrow \qq  T > {T_0\, N \over (w+1) \ln N} \ ,
\ee
the fully symmetric irrep is unstable, the stable one being a "modified symmetric" configuration with a small number
$s \ll n\sim N^2$ of boxes in lower rows. We call this a "pseudo-phase transition," as all thermodynamic quantities
are continuous across $\tau\bb=\bb1/(w\bb+\bb1)$ 
to leading order in $N$.
For the distribution $\rho$, this corresponds to $\rho(x)$ deviating slightly from the step
function form \eqn{ground} near $x=1$, developing a smooth transition from $\rho =1$ to $\rho=0$.
The exact shape of this transition may be of mathematical interest, but it is irrelevant for the thermodynamics of the model. 

\begin{figure} [th!]
\begin{center}
\includegraphics[height= 7.5 cm, angle=0]{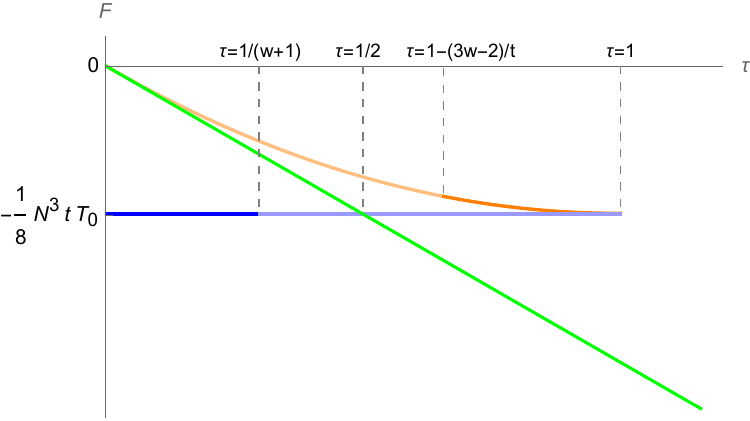}
\end{center}
\vskip -0.75 cm
\caption{\small{The free energy of the lumped (green), lumped-disjoint (orange), and disjoint (blue) phases as a function of the scaled temperature $\tau$
 and for $t> 3w-2$. For $\tau<1/(w+1)$ the disjoint phase is the symmetric irrep (deep blue),
while for $1/(w+1)<\tau<1$ it has a subleading number of boxes in lower rows (light blue). Likewise, for
$\tau<1-(3w-2)/t$ the lumped part of the lumped-disjoint phase is in the dilute configuration (light orange), while for
$1-(3w-2)/t <\tau <1$ it is in the dense configuration (deep orange). We have depicted the case with $t>2 (3 w-2)$. For $3 w-2<t<2 (3 w-2)$ the second 
and third dashed vertical lines are interchanged. }}
\label{LargeTfree}
\end{figure}

Similarly, by using the results of \cite{Polychronakos:2023qwz}, or by referring to fig. \ref{PhaseDiag},
if the parameter $t$ of the model is below its large-temperature critical
value, i.e.  $t<3w-2$, then the lumped part $\rho$ in the lumped-disjoint phase will be in the dense configuration for all $\tau$. 
If, however, $t>3w-2$, then for
${\tilde t} < 3w -2$ the lumped part will be in the  dense phase, while for ${\tilde t}>3w-2$ it will be in the dilute configuration; that is,
\be
\begin{split}
&\text{dilute-disjoint:}\quad 0< \tau < 1-{3w-2 \over t}\ ,
\\
&\text{dense-disjoint:} \quad 1-{3w-2 \over t} <\tau <1\ .
\end{split}
\ee
Nevertheless, the two configurations have the same free energy to leading order in $N$ as a function of $\tau$,
so they do not constitute thermodynamically  distinct phases and this is another pseudo-phase transition.
The free energies of all phases at large temperature as a function of $\tau$, and for $t> 3w-2$, are depicted in figure \ref{LargeTfree}.
For $t< 3w-2$ the figure is similar, with the lumped part of the metastable mixed phase being entirely in the
dense configuration (the entire curve is in deep orange and the line at $\tau=1-(3w-2)/t$ missing.

\no
The above results can be compared to those of \cite{Phases}, which examined the case $w=2$ and finite $N$.
The transition between the stable and metastable phases involved three temperatures $T_0<T_1<T_c$: for 
$T<T_0$ the ferromagnetic phase was stable and the paramagnetic one unstable; for $T_0<T<T_1$ the paramagnetic
phase became metastable; for $T_1<T<T_c$ the ferromagnetic phase turned metastable and the  paramagnetic one
became stable; and for $T>T_c$ the ferromagnetic phase ceased to exist and the paramagnetic one remained stable.
For large $N$, the temperatures $T_1$ and $T_c$ were found to behave as $T_1\simeq T_c/2 \simeq {N\over \ln N} T_0$.
These precisely match the values $\tau=\ha $ and $\tau=1$ in \eqn{Ttau} which mark the temperatures for similar phase
transitions in our case. The modified structure of these phases, however, as well as the
existence of the metastable paramagnetic-ferromagnetic phase, were lost in the finite-$N$ analysis.

\begin{figure} [th!]
\begin{center}
\includegraphics[height= 7.5 cm, angle=0]{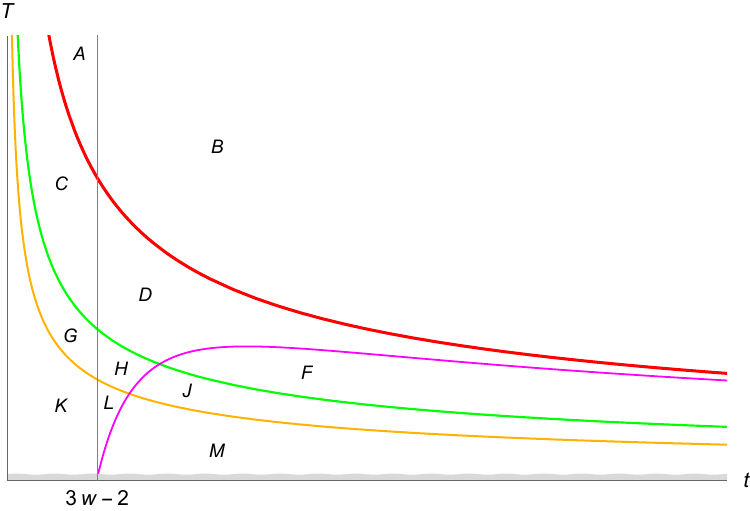}
\end{center}
\vskip -.5 cm
\caption{\small{The phase diagram for temperatures $T\bb \sim\bb T_L$ at fixed number of
atoms $n$. The upper (red) curve $T=T_L /\sqrt t$ marks a true phase transition and the middle (green)
curve $T = T_L/2\sqrt t$ marks a metastability transition, while the remaining (orange, purple, gray) curves represent
pseudo-transitions. The various phases are:\\
dense-lumped:$\,$A,C$\,$(stable)$\,$G,K$\,$(metastable);
dilute-lumped:$\,$B,D,F$\,$(stable)$\,$H,J,L,M$\,$(metastable)\hfill\\
modified symmetric: C,D,F (metastable) G,H,J (stable);
fully symmetric: K,L,M (stable)\\
dense-disjoint: C,D,G,H,K,L (metastable);
dilute-disjoint: F,J,M (metastable)\\
The shaded region near the $t$ axis represents temperatures $\sim\bb T_0$ with phases depicted in fig.
\ref{PhaseDiag}}}
\label{LargeTphases}
\end{figure}

The large-temperature phases on the $t\! -\! T$ plane can be depicted by trading $N$
for the parameter $t$ and the "volume" (number of atoms) $n$, and keeping $n$ constant as we vary $t$ and $T$.
In this parametrization, large temperatures \eqn{Ttau} become
\be
T = T_L {\tau \over \sqrt t} ~,~~~\text{with}~~~ T_L = 4 T_0\, {\sqrt n \over \ln n}\ ,
\ee
where we used $\ln n \gg 1, \ln t$. Then the curve corresponding to $\tau =1$ marks a true phase transition and the
one corresponding to $\tau = \half$ a metastability transition. The curves for $\tau = 1/(3w-2)$, $t=3w-2$,
and $\tau = 1-(3w-2)/t$ separate phases with qualitatively different features but thermodynamically represent  pseudo-transitions,
as explained earlier. The full phase diagram is presented in fig. \ref{LargeTphases}.

\section{Conclusions}
\label{concl}

The phase structure of the $SU(N)$ ferromagnet at large $N$ manifests qualitatively new properties as compared to
the finite-$N$ case. The main novel features of the large-$N$ model are, first, the emergence of two distinct
temperature scales, and second, the appearance of additional phases and sub-phases.
The two temperature scales are related by a scale factor of order $N\!/\!\ln N$, the lower one being relevant to the stability
of the paramagnetic phases and the higher one to the stability of the ferromagnetic phases. The appearance
of additional phases occurs both within the paramagnetic regime, which splits into two phases separated by
a phase transition and leads to a triple point, and the ferromagnetic phase, which can exhibit small deviations
from the fully symmetric irrep of the finite-$N$ case, but also develops a second, thermodynamically distinct
metastable phase that deviates substantially from the symmetric irrep.

The element driving the emergence of new features in the model is the Vandermonde term in the effective action
(\ref{Zmw}, \ref{mww}), which is thermodynamically suppressed for finite $N$ but becomes relevant at large $N$.
This leads to a paramagnetic phase that corresponds to an irrep of $SU(N)$ other than the singlet, which changes
with the temperature, and to generalized symmetry breaking patterns in the ferromagnetic phase.

Although the paramagnetic phases are metastable at low temperatures, and the novel mixed
paramagnetic-ferromagnetic phase is metastable at all temperatures, their presence is physically significant.
By Arrhenius' law, the transition of a metastable state to a fully stable one, when driven only by thermal fluctuations,
is exponentially suppressed and the transition time is exceedingly large. For all practical purposes, metastable
states are stable if left unperturbed, and only external perturbations (impurities, shaking the system etc.) can induce
their decay. Our results, therefore, can be physically relevant in the appropriate context.

Potential applications of the large-$N$, general-$w$ model studied in this work extend to any system where
the number of degrees of freedom per atom $N$ grows large and becomes comparable to the square root of the
number of atoms $\sqrt n$, thus making the analysis in which the Vandermonde terms is neglected unreliable.
Even for moderately high $N$, the large-$N$ results may better approximate the thermodynamics of the model
than the finite-$N$ results, in cases where the two differ. Such situations can arise either when $N$ is
relatively high, or when the number of atoms $n$ is not very high, which is the case in some experimentally
prepared systems (for instance, in lattices created with trapped ions, the number of sites is of order a few tens
 \cite{Korenblit:2012fqk}).

The model could also be relevant to more exotic situations, such as the physics of the quark-gluon plasma (for a review see, e.g.,
\cite{Pasechnik:2016wkt}),
that is, a fluid of particles carrying color degrees of freedom which are assumed to ferromagnetically interact.

An obvious next step in further analyzing the large-$N$ model would be the addition of external non-Abelian
magnetic fields. Their inclusion in the finite-$N$ model led to a highly nontrivial phase diagram in the
temperature-magentic field plane, with several (meta)stable states and phase transitions appearing.
In the large-$N$ case, the thermodynamic phase space would be enlarged into the three-dimensional
$T\!-\!t\!-\!B$ space, $B$ representing the magnetic field in a single direction of $SU(N)$, potentially leading to a
very rich phase structure. 

In addition, several possible generalizations of our model exist, similar to the ones open in the
finite-$N$ case. A system of atoms each carrying an irrep of $SU(N)$ other than the fundamental, representing a
case with reduced symmetry among the states of the atom, would be interesting to consider. Alternatively,
systems involving 3-atom or higher interactions, which would manifest in the appearance of higher Casimirs in
the energy term, would be interesting to analyze in the large-$N$ limit.

Finally, applications in matrix models and large-$N$ Yang-Mills theories could also be envisaged
(see, e.g. \cite{Mar,Tur,Romiti:2023hbd} and \cite{Gross:1980he,Wadia:1980cp,Russo:2020eif,Russo:2020pnv}). 
The possible relevance of the ferromagnetic term in the physics of
microstates in two-dimensional black holes \cite{Witten:1991yr, Mandal:1991tz} as studied in 
\cite{Kazakov:2000pm,Kazakov:2001pj} and more recently in \cite{Betzios:2022pji, Ahmadain:2022gfw},
as well as in the deconfinement/Hagedorn transition in large-$N$ gauge theories
\cite{Bo,AMMPR,HaMaSu,Hanada,ark} would be an interesting and important issue.
This and other related questions merit further investigation.

\subsection*{Acknowledgements}

The research of A.P. was supported by the National Science Foundation 
under grant NSF-PHY-2112729 and  by PSC-CUNY grants 65109-00 53 and 6D136-00 02.\\
K.S. would like to thank the Department of Theoretical Physics at CERN for hospitality and financial support during the a stage of this research.


\vskip -1.5cm
$ $

\end{document}